\documentclass[%
 aip,
 jmp,%
 amsmath,amssymb,
 reprint,%
]{revtex4-1}
\usepackage{appendix}
\usepackage{graphicx}% Include figure files
\usepackage{dcolumn}% Align table columns on decimal point
\usepackage{bm}% bold math
\usepackage{color}

\begin{document}

%\preprint{AIP/123-QED}

\title[Sample title]{Aging Feynman-Kac Equation}% Force line breaks with \\
%\thanks{Footnote to title of article.}

\author{Wanli Wang}
 %\altaffiliation[Also at ]{Physics Department, XYZ University.}%Lines break automatically or can be forced with \\
\author{Weihua Deng}%
 \email{dengwh@lzu.edu.cn}
\affiliation{
 $^1$School of Mathematics and Statistics, Gansu Key Laboratory of Applied Mathematics and Complex
Systems, Lanzhou University, Lanzhou 730000,  P.R. China%\\This line break forced with \textbackslash\textbackslash
}%

%\author{C. Author}
% \homepage{http://www.Second.institution.edu/~Charlie.Author.}
%\affiliation{%
%Second institution and/or address%\\This line break forced% with \\
%}%

\date{\today}% It is always \today, today,
             %  but any date may be explicitly specified

\begin{abstract}
Aging, the process of growing old or maturing, is one of the most widely seen natural phenomena
in the world. For the stochastic processes, sometimes the influence of aging can not be ignored.
For example, in this paper, by analyzing the functional distribution of the trajectories of aging
particles performing anomalous diffusion, we reveal that for the fraction of the occupation time
$T_+/t$ of strong aging particles, $\langle (T^+(t)^2)\rangle=\frac{1}{2}t^2$ with coefficient
$\frac{1}{2}$, having no relation with the aging time $t_a$ and $\alpha$ and being completely
different from the case of weak (none) aging. In fact, we first build the models governing the
corresponding functional distributions, i.e., the aging forward and backward Feynman-Kac
equations; the above result is one of the applications of the models. Another application of the
models is to solve the asymptotic behaviors of the distribution of the first passage time,
$g(t_a,t)$. The striking discovery is that for weakly aging systems, $g(t_a,t)\sim
t_a^{\frac{\alpha}{2}}t^{-1-\frac{\alpha}{2}}$, while for strongly aging systems, $g(t_a,t)$
behaves as $ t_a^{\alpha-1}t^{-\alpha}$.
\end{abstract}

\pacs{02. 50. -r,  05. 30. Pr,  05. 40. -a,  05. 10. Gg}% PACS, the Physics and Astronomy
                             % Classification Scheme.
%\pacs{02. 50. -r,  05. 30. Pr,  05. 40. -a,  05. 10. Gg }% PACS,  the Physics and Astronomy
%   02. 50. -r   Probability theory, stochastic processes, and statistics
%  05.30.Pr	Fractional statistics systems (anyons, etc.)

% 05.40.-a Fluctuation phenomena, random processes, noise, and Brownian motion (for fluctuations in superconductivity, see 74.40.-n; for statistical theory and fluctuations in nuclear reactions, see 24.60.-k; for fluctuations in plasma, see 52.25.Gj; for nonlinear dynamics and chaos, see 05.45.-a)
%
%¡¡05.10.Gg Stochastic analysis methods (Fokker-Planck, Langevin, etc.)

% 02.50.Ey	Stochastic processes
%  45.10.Hj	Perturbation and fractional calculus methods

%Use showkeys class option if keyword
                              %display desired
\maketitle

%\begin{quotation}
%The ``lead paragraph'' is encapsulated with the \LaTeX\
%\verb+quotation+ environment and is formatted as a single paragraph before the first section heading.
%(The \verb+quotation+ environment reverts to its usual meaning after the first sectioning command.)
%Note that numbered references are allowed in the lead paragraph.
%%
%The lead paragraph will only be found in an article being prepared for the journal \textit{Chaos}.
%\end{quotation}

\section{Introduction}\label{sect1}
One of the most omnipresent phenomena in nature is aging, being clearly a process of time $t$.
The models naturally coming into our mind are the stochastic processes, the leading examples of which are the renewal processes. As the generalization of Poisson process \cite{Papoulis:84}, in this paper, we mainly focus on the renewal processes with independent identically distributed (i.i.d.) holding times between any adjacent two renews, and the distribution of holding times is power law with divergent first moment. Not like Poisson process, in this case, the power law renewal process is not longer Markovian, which leads to nonstationarity, exhibiting aging behaviors
\cite{Barkai2002Aging}.

For the power law renewal process, if there is a jump for each renewal, we get the compound power
law renewal process; the size of jump is generally an i.i.d. random variable with specified
probability distribution. This compound renew process can effectively characterize anomalous diffusion \cite{Krusemann:01}, whose mean squared displacement is a nonlinear function of time $t$, i.e.,
\begin{equation}\label{anomalousdiffusion}
\left< x^2(t) \right> \simeq K_\alpha t^\alpha;
\end{equation}
if the second moment of the variable of the size of the jump is bounded, the process describes subdiffusion, which means $\alpha \in (0,1)$; on the other side, if the second moment diverges,
the process is for the competition between long rests and long jumps \cite{Metzler:00}, and the resulted macroscopic behavior may be superdiffusion, subdiffusion, or even normal diffusion.

%If there are jumps
%
%The compound renewal processes with power law distribution of holding times
%
% can effectively
%
%
%Krusemann:01
%
%
%Today, the term anomalous diffusion
%in statistical physics, biological physics, and geophysics refers to processes, whose mean
%squared displacement is no longer linear in time
%
%
%Anomalous diffusion is a diffusion process with non-linear relationship to time $t$, i.e, $\langle r^2(t)\rangle\sim t^\gamma $ and $\gamma\neq 1$, in contrast to  the normal diffusion process, in which the index of the mean squared displacement is $\gamma=1$.

Anomalous diffusion phenomena are ubiquitous in real life. For example, they are found in systems including ultra-cold atoms, telomeres in the nucleus of cells \cite{Bronstein2009Transient}, bacterial motion \cite{Nossal1983Stochastic}, and even the flight of an albatross~\cite{Viswanathan1996levy}. Anomalous diffusion in the presence or absence of force fields has been modelled in many ways  such as generalised diffusion equations~\cite{OShaughnessy1985Analytical}, fractional Brownian motion~\cite{Mandelbrot1968Fractional}, continue time random walk~\cite{Klafter2011First,Metzler20001The} (CTRW), Langenvin equations \cite{Bruce1982Linear,Fogedby1994Levy}, and so on. The compound power law renewal process is a specific class of CTRW model, where both waiting time and jump length are i.i.d. random variables, in particular, the distribution of waiting time is power law with divergent first moment.
Generally, for the CTRW model, the physical clock is assumed to start at the first step of the process, i.e., we have to observe or measure the system immediately after its preparation in the present state. For the real physical process, the observation time should not be exactly the starting time of the process. Monthus and Bouchaud~\cite{Monthus1996Models} introduce a CTRW framework, which can be used to study aging behaviors. It is called generalized CTRW or aging continuous time random walk (ACTRW) \cite{Barkai2002Aging}, where the measure time starts not at time $t=0$ but at some later instant time $t=t_a>0$ and $t_a$ denotes the aging time.

There are already some research works for the aging in complex systems \cite{Barkai2003Aging,Klafter2011First,Marinari1993On,Doussal1999Random,Sokolov2001Linear}; and the aging behaviors are observed in the fluorescence of single nanocrystals \cite{Brokmann2003Statistic}, polymers \cite{Struick1978Physical}, time average of single particle trajectories in scale-free anomalous diffusion processes \cite{Schulz2013Aging}. Here we further discuss the aging behavior in complex system by analyzing the functional distribution of the trajectories of the aging particles.
%For  CTRW model, the physical clock is started at the first step of the process, i.e., we have to observe or measure the system immediately after its preparation in the present state. In fact, sometimes it's not easy to observe of the process immediately after it stared. Monthus and Bouchaud~\cite{Monthus1996Models} introduced a CTRW framework, which shows aging behaviors. We call this generalized CTRW  or aging continuous time random walk (ACTRW). Aging continue time random walk is a model to observe such a process, i.e., the measure time starts not at time $t=0$ but at some later instant time $t=t_a>0$ and $t_a$ denotes as aging time. Aging phenomena \cite{Barkai2003Aging,Klafter2011First,Marinari1993On,Doussal1999Random,Sokolov2001Linear} are observed in a large number of systems, and they are used as a tool to research complex systems. Furthermore, aging in diffusion process yields many interesting properties on dynamic in disordered medium. The complex dynamical systems illustrating aging behavior are quite extensive, including the fluorescence of single nanocrystals \cite{Brokmann2003Statistic}, polymers \cite{Struick1978Physical}, time average of single particle trajectories in scale-free anomalous diffusion processes \cite{Schulz2013Aging}.
The functional $A$ is defined as
\begin{equation}\label{31eq01}
 A=\int_0^t U(x(\tau))d\tau,
\end{equation}
where $U$ is a prescribed function and $x(t)$ is the trajectory of the particle. Based on Eq. (\ref{31eq01}), many interesting applications have been dug out. If one is concerned about the time spent by a particle in a given domain, it can be made by taking $U=1$ if the particle lies in the domain and $0$ otherwise \cite{Luchinin2014Application,Bar-Haim1998On,Wu2016Tempered}. The functional is also used to study NMR \cite{Grebenkov2007NMR}; under the influence of inhomogeneous magnetic $U(x(t))$, the total phase accumulated along the trajectory of a nucleus during the time from $0$ to $t$ is taken as
\begin{equation*}\label{31eq02}
\vartheta(t)=\gamma\int_0^t U(x(\tau))d\tau,
\end{equation*}
where $\gamma$ is the nuclear gyromagnetic ratio; and $U(x)$ is respectively specified as $x$ and $x^2$ to calculate the macroscopic measured signal $E=\langle \exp(i\vartheta(t))\rangle$; then, NMR indirectly encodes information regarding the trajectory of the particles. If $U=\delta(x-a)$, the functional $A$ denotes the local time at the fixed level $a$, which is an important quality in probability \cite{Pitman1999The}. Yet another instance used in finance is the case $U(x)=\exp(x)$, which illustrates the price of an Asian stock option in the Black-Scholes framework \cite{Fischer1973The}. Interestingly, these  functionals $A$ and their related variants are also used as a powerful tool in mathematics and physics, i.e.,  Feynman-Kac formula, which allows to study the functionals in a quantum mechanical framework \cite{Majumdar2005Brownian,Kac1949On,Perret2015On,Baule2006Investigation,Majumdar2002Large,Carmi2010On}.

This paper is organized as follows. In Sect.~\ref{sect2}, based on the ACTRW model, we derive the forward Feynman-Kac equation with (tempered) power law waiting time. We start from the simple case, i.e.,  the probability density function (PDF) of the step length is taken as $f(x)=\frac{1}{2}[\delta(x-a)+\delta(x+a)]$, which implies that the particles can only move to the left or right direction with the same probability; then we use Gaussian distribution and power law distribution, respectively, as the PDF of jump length.
%; finally, we consider the tempered power law waiting, and derive its corresponding equation.
If letting $p=0$, one obtains a generalization of the Montroll-Weiss equation for ACTRW, which agrees with the previous result \cite{Barkai2003Aging,Klafter2011First}. In Sect.~\ref{sect22}, we obtain the corresponding backward Feynman-Kac equations. Based on the derived equations, some applications are presented, such as the occupation in half space $T^{+}$, the moments of $T^{+}$ and $(T^{+})^2$. The behaviors of $\langle(T^{+})^2\rangle$ are different for strong and weak (none) aging. Furthermore, the asymptotic behaviors of the first passage time are analyzed for both the cases of strong and weak aging. In the last section, we conclude the paper with some discussions.

\section{Derivation of the fractional forward aging Feynman-Kac Equation}\label{sect2}
Based on ACTRW, we derive its corresponding forward Feynman-Kac equations, including the cases that the particles move with constant jump length, and the power law jump length and (tempered) power law waiting times.
%For simplify, we consider the particles moving on an infinite one-dimensional lattice at constant  step length; furthermore, Both the power law and tempered power law waiting time are discussed.
Now, we first briefly outline the main ingredients of the CTRW \cite{Montroll1965Random,Kenkre1973Generalized,Gorenflo2007Continue} and ACTRW \cite{Barkai2003Aging,Klafter2011First,Marinari1993On,Brokmann2003Statistic,Struick1978Physical}.   CTRW and ACTRW are defined as follows: a walker is trapped on the origin for the time $t_{1}$, then makes a jump and the displacement is $x_{1}$; the walker is further trapped on $x_1$ for time $t_2$, and then jumps to a new position; this process is then renewed.
Both of them are characterized by a set of waiting times $\{t_1, t_2, \cdots, t_n, \cdots\}$
 and the displacements $\{x_1, x_2, \cdots, x_n, \cdots\}$. For both CTRW and ACTRW, all $x_{i}$ are i.i.d. with a common PDF $f(x) \sim |x|^{-1-\mu}$ or Gaussian distribution. For CTRW, all $t_i$ are i.i.d. with a common PDF $\phi(t)$ and we start to observe the process from the origin $t=0$. While for ACTRW, observation time  starts not at time $t=0$ but at later instant of time $t=t_a>0$, and the time $t_a$ is called the aging time. The ACTRW modifies the statistic of the time for the first jump, then we have to know how long the walk has to wait from its initiation time until making its first step after the observation stated, i.e., the waiting time PDF of the first jump, which is denoted by $\omega(t_a, t)$. While all the other $t_i\, (i>1)$ are i.i.d. with a common PDF $\phi(t)$. If $t_a=0$, we have $\omega(t_a, t)=\phi(t)$. Then ACTRW model reduces to CTRW one. We suppose that the walk's position at $t=t_a$ is $\tilde{x}(t_a)$; from the  time on (a new starting point), what we are interested in is its position at time $t$, i.e.,
 \begin{equation}\label{33eq07}
  % x(t)=\tilde{x}(t+t_a)-\tilde{x}(t_a).
  x(t)=\tilde{x}(t+t_a).
 \end{equation}
 %Supposing the waiting time between each step is $\phi(t)$, we
Consider the broad distribution characterized by a power law heavy tail \cite{Klafter2011First,Godreche2001Statistics}
\begin{equation}\label{33eq01}
  \phi(t) \sim \frac{\alpha}{\Gamma(1-\alpha)}\frac{\tau^\alpha}{t^{1+\alpha}},
\end{equation}
where $\tau$ is a microscopic time scale and the index $\alpha>0$. Note that the moments of $\phi(t)$ diverge for $\alpha<1$, while the first moment $\langle \tau\rangle$ is finite but the higher moments are divergent for $1<\alpha<2$. If $\alpha>2$, both the first and the second moments are bounded. Using the Tauberian theorem \cite{Wiley1971An,Klafter2011First} from Eq. (\ref{33eq02}), in Laplace space, we can obtain
\begin{equation}\label{33eq03}
\mathcal{L}[\phi(t)]=\phi(s) \sim 1- B_\alpha s^\alpha,
\end{equation}
where $s$ is conjugate to $t$, $B_\alpha=\tau^\alpha$, and $0<\alpha<1$.
%For $1<\alpha<2$, we have $\phi(s) \sim 1- \langle \tau \rangle s + as^\alpha$. if $\phi(t)$ is a narrow distribution, performing Laplace transform, we have
%\begin{equation*}\label{33eq04}
%\hat{\psi}(s) \sim 1- \langle \tau \rangle s + \frac{1}{2}\langle \tau^2 \rangle u^2,
%\end{equation*}

 We denote $G(x, A, t, t_a)$ as the joint PDF of $x$ and $A$ at time $t$ with aging time $t_a$, where $A$ is the functional defined as $A=\int_0^tU(x(\tau))d\tau$. The difference between CTRW and ACTRW is the first step, i.e., the distribution of the waiting time of the first step is different, which plays an important role in the process of the derivation of the  aging Feynman-Kac equation. Let $Q_1(x, A, t, t_a)$ be  the joint PDF of $x$ and $A$ at time $t$ with aging time $t_a$ for the first step. $Q_0(x, A, t)$ is the joint PDF of  $x$ and $A$ at the starting observation time $t$.

\subsection{Discrete step length PDF}\label{sect21}
For simplicity, we first consider a particle walking on an infinite one-dimensional lattice and the length of each  is a constant $a$. The particle is just allowed to jump to its nearest positions with the same probability to the left or right direction. Using the definition of ACTRW yields
\begin{equation}\label{33eq11}
\begin{split}
  Q_1(x,& A, t, t_a)  =\int_0^t \frac{1}{2}\omega(t_a, \tau)\\
  &\Big(Q_0(x+a, A-\tau U(x+a), t-\tau) \\
  &+Q_0(x-a, A-\tau U(x-a), t-\tau) \Big)d\tau,
\end{split}
\end{equation}
where $Q_0(x,A,t)$ is the initial distribution, %i.e., no jump is made,
and $\omega(t_a, \tau )$ is the forward waiting time PDF. Using Laplace transform \cite{Dyke2014An,Oberhettinger1973Tables,Erdelyi1954Tables} with respect to $A$, i.e., $A\rightarrow p$
\begin{equation}\label{33eq12}
\begin{split}
 Q_1(x, &p, t, t_a)=\int_0^t \frac{1}{2}\omega(t_a,\tau)\\
 &\cdot\Big(\exp(-p\tau U(x+a))Q_0(x+a, p, t-\tau) \\
 &+\exp(-p\tau U(x-a))Q_0(x-a, p, t-\tau)  \Big)d\tau.
\end{split}
\end{equation}
Taking Laplace transform ($t\rightarrow s$ ) and using the shift property of Fourier transform \cite{Oberhettinger1973Tables,Erdelyi1954Tables} ($x\rightarrow k$ ), we have
\begin{equation}\label{33eq13}
 Q_1(k, p, s, t_a)  =\cos (k a)\omega\Big(t_a, s+pU\Big(-i\frac{\partial}{\partial k}\Big)\Big)Q_0(k,p,s),
\end{equation}
where we use the relation $\mathcal{F}[g(x)f(x)]=g(-i\frac{\partial}{\partial k})f(k)$ and denote $f(k)=\mathcal{F}[f(x)]$.
%which can be confirmed by Taylor's formula and the property of  Fourier transform.
By the similar way as above, for $n\geqslant 1$, there exists
\begin{equation}\label{33eq14}
\begin{split}
  Q_{n+1}(x,& A, t, t_a)  =\int_0^t \phi(\tau)\frac{1}{2}\\
  &\cdot\Big(Q_n(x+a, A-\tau U(x+a), t-\tau, t_a)+\\
  &Q_n(x-a, A-\tau U(x-a), t-\tau, t_a) \Big)d\tau,
\end{split}
\end{equation}
where $\phi(\tau)$ is the PDF of waiting time between jumps for $n_{th} (n\geqslant1)$  step.
Using double Laplace transform ($A\rightarrow p$ and $t\rightarrow s$) and Fourier transform ($x\rightarrow k$) we can obtain
\begin{equation}\label{33eq15}
 Q_{n+1}(k, p, s, t_a)  =\cos (k a) \phi\Big(s+pU\Big(-i\frac{\partial}{\partial k}\Big)\Big)Q_n(k,p,s, t_a)
\end{equation}
with $n\geq1$, which results in
\begin{equation}\label{33eq16}
\begin{split}
  ~ & \sum_{n=1}^{\infty}Q_{n}(k, p, s, t_a)=\frac{Q_1(k, p, s, t_a)}{1-\cos (ka) \phi(s+pU(-i\frac{\partial}{\partial k}))}\\
    & =\frac{\cos (ka) \omega(t_a,s+pU(-i\frac{\partial}{\partial k})) Q_0(k, p, s)}{1-\cos (ka) \phi(s+pU(-i\frac{\partial}{\partial k}))}.
\end{split}
\end{equation}
We can now return to our random walk and the joint  PDF of a walk at time $t$ with aging time $t_a$ is given by
\begin{equation}\label{33eq21}
\begin{split}
  G(x,A,t,t_a) & =\Big(1-\int_0^t\omega(t_a,\tau)d\tau\Big)\delta(A-tU(x))\delta(x-x_0)\\
    &+ \int_0^t\Phi(\tau)\sum_{n=1}^{\infty}Q_n(x, A-\tau U(x), t-\tau, t_a) d\tau,
\end{split}
\end{equation}
where $x_0$ is the initial position, i.e., $x_0=x(t=0)$ and $\Phi(t)$ is  the survival probability \cite{Klafter2011First,Krusemann2014First} on a site, i.e.,  the probability that the waiting time on a site exceeds t
\begin{equation}\label{33eq02}
  \Phi(t)=\int_{t}^{\infty} \phi(\tau) d\tau=1-\int_{0}^{t} \phi(\tau) d\tau.
\end{equation}
In Laplace space, $ \Phi(s)=\frac{1-\phi(s)}{s}$. Performing Laplace transform from $A$ to $p$ leads to
\begin{equation}\label{33eq22}
\begin{split}
  G(&x,p,t,t_a)  =\Big(1-\int_0^t\omega(t_a,\tau)\Big)\exp(-tpU(x))\delta(x-x_0) \\ &+\int_0^t\Phi(\tau)\sum_{n=1}^{\infty}\exp(-\tau pU(x))Q_n(x, p, t-\tau, t_a) d\tau.
\end{split}
\end{equation}
In Laplace space,
\begin{equation}\label{33eq23}
\begin{split}
  G(x,p,s,t_a) &  =\frac{1-\omega(t_a, s+pU(x))}{s+pU(x)}\delta(x-x_0)\\
  &+\Phi(s+pU(x))\sum_{n=1}^{\infty}Q_n(x, p, s, t_a).
\end{split}
\end{equation}
Taking Fourier transform on the above equation yields
\begin{equation}\label{33eq24}
\begin{split}
  G(k,p,s,t_a) & =\frac{1-\omega(t_a, s+pU(-i\frac{\partial}{\partial k}))}{s+pU(-i\frac{\partial}{\partial k})}\exp(ik x_0)\\
  &+\frac{1-\phi(s+pU(-i\frac{\partial}{\partial k}))}{s+pU(-i\frac{\partial}{\partial k})}\cdot\\
  &\frac{\cos(ka)\omega(t_a,s+pU(-i\frac{\partial}{\partial k}))Q_0(k,p,s)}{1-\cos(ka)\phi(s+pU(-i\frac{\partial}{\partial k}))}.
\end{split}
\end{equation}
Note that Eq.~\eqref{33eq24} is valid for all kinds of PDF of waiting time. Omitting the singular part of Eq.~(\ref{33eq24}), i.e., the unmoving part, and performing Laplace transform with  respect to $t_a$, $t_a\rightarrow u$, result in
 \begin{equation}\label{33eq25}
\begin{split}
  G(k,p,s,u)
  &=\frac{1-\phi(s+pU(-i\frac{\partial}{\partial k}))}{s+pU(-i\frac{\partial}{\partial k}))}\\
  &\frac{\cos(ka)\omega(u,s+pU(-i\frac{\partial}{\partial k}))Q_0(k,p,s)}{1-\cos(ka)\phi(s+pU(-i\frac{\partial}{\partial k}))}.
\end{split}
\end{equation}
Using  previous work given by Godr\`{e}che and Luck \cite{Godreche2001Statistics}, in Laplace space,  the forward waiting time PDF of $\omega(t_a,t)$  can be given by
\begin{equation}\label{33eq26}
  \omega(u,s)=\frac{1}{1-\phi(u)}\frac{\phi(u)-\phi(s)}{s-u}.
\end{equation}
Based on Eq.~\eqref{33eq25} and Eq.~\eqref{33eq26}, we study special case of $\phi(t)$ and obtain its corresponding forward equation. Consider broad distribution of waiting times with index $\alpha<1$. Taking the limit $k\rightarrow 0$,  substituting Eq. (\ref{33eq03}) into Eq. (\ref{33eq24}), and expanding $\cos(ka)$ as series in $k$ (i.e., $\cos(ka)\sim 1-k^2a^2/2$), we get
% \begin{equation}\label{33eq41}
%\begin{split}
%  sG(k,p,s,u) & =-pU(-i\frac{\partial}{\partial k})G(k,p,s,u)-\frac{(s+pU(-i\frac{\partial}{\partial k}))^{1-\alpha}k^2}{2B_a}G(k,p,s,u)+\omega(u,s+pU(-i\frac{\partial}{\partial k}))Q_0(k,p,s)\\
%  &+\frac{1/u-\omega(u, s+pU(-i\frac{\partial}{\partial k})))}{B_a(s+pU(-i\frac{\partial}{\partial k}))^{\alpha}}(1-\cos(ka)+B_a(s+pU(-i\frac{\partial}{\partial k})^\alpha\cos(ka))\exp(ik x_0),
%\end{split}
%\end{equation}
  \begin{equation}\label{33eq41}
\begin{split}
  sG(k,p,s,u) & =-pU\left(-i\frac{\partial}{\partial k}\right)G(k,p,s,u)\\
  &-a^2\frac{k^2(s+pU(-i\frac{\partial}{\partial k}))^{1-\alpha}}{2B_\alpha}G(k,p,s,u)\\
  &+\omega\left(u,s+pU\left(-i\frac{\partial}{\partial k}\right)\right)Q_0(k,p,s).\\
\end{split}
\end{equation}
Supposing that the initial distribution $Q_0(x,A,t=0)=\delta(t)\delta(x-x_0)\delta(A)$,
%, by the similar way, we can obtain
%\begin{equation}\label{33eq45}
% \begin{split}
%     \mathcal{F}^{-1}\mathcal{L}^{-1}\mathcal{L}^{-1}&\omega\Big(u,s+pU\Big(-i\frac{\partial}{\partial k}\Big)\Big)Q_0(k,p,s)\\
%     &=\exp(-tpU(x))\omega(t_a,t)\delta(x-x_0).
% \end{split}
%\end{equation}
%in the following, using the relation Eq. (\ref{33eq26}), we consider the last line of Eq. (\ref{33eq41}), which is divided into two parts
%setting the initial position $Q_0(x,A,t=0)=\delta(t)\delta(x-x_0)\delta(A)$,
using the above formulas, and performing the inverse transform, we get
%\begin{equation}\label{33eq53}
%\begin{split}
% \frac{\partial}{\partial t}G(x, p,t,t_a) & =\frac{a^2}{2B_a}\frac{\partial}{\partial x^2} ~_{0}D_t^{1-\alpha}G(x, p,t,t_a)-pU(x)G(x, p,t,t_a)+\exp(-tpU(x))\omega(t_a,t)\delta(x) \\
%    & -\delta(x-x_0)\delta(t)\delta(t_a)+n(x,p,t,t_a)+m(x,p,t,t_a),
%\end{split}
%\end{equation}
\begin{equation}\label{33eq53}
\begin{split}
 \frac{\partial}{\partial t}G(x, p,t,t_a) & =\frac{a^2}{2B_\alpha}\frac{\partial^2}{\partial x^2}D_{t}^{1-\alpha}G(x, p,t,t_a)\\
 &-pU(x)G(x, p,t,t_a)-\delta(t)\\
 &+\exp(-tpU(x))\omega(t_a,t)\delta(x-x_0),
\end{split}
\end{equation}
where $D_{t}^{1-\alpha}$ is the fractional substantial derivative \cite{Wu2016Tempered}, being  defined by
 \begin{equation}\label{33eq42}
 \begin{split}
  D_{t}^{1-\alpha}&G(x,p,t,t_a)=\frac{1}{\Gamma(\alpha)}\Big[\frac{\partial}{\partial t}+pU(x)\Big]\\
  &\cdot\int_0^t\frac{\exp(-(t-\tau)pU(x))}{(t-\tau)^{1-\alpha}}G(x,p,\tau,t_a)d\tau;
  \end{split}
\end{equation}
%$n(x,p,t,t_a)$ and $m(x,p,t,t_a)$ are the  motionless part of Eq. (\ref{33eq25}).
%\begin{equation}\label{33eq31}
%  ~_{0}D_t^{1-\alpha}G(x,p,t,t_a)=\frac{1}{\Gamma(\alpha)}[\frac{\partial}{\partial t}+pU(x)]\int_0^t\frac{\exp(-(t-\tau)pU(x))}{(t-\tau)^{1-\alpha}}G(x,p,t,ta)d\tau,
%\end{equation}
%using the definition of convolution and shift theorem of Laplace transform we have
%\begin{equation}\label{33eq32}
%  ~_{0}D_t^{1-\alpha}G(x,p,s,t_a)=[s+pU(x)]^{1-\alpha}G(x,p,s,t_a),
%\end{equation}
%from the definition of $~_{0}D_t^{1-\alpha}G(x,p,t,t_a)$, it can be noticed that the operate $~_{0}D_t^{1-\alpha}G(x,p,t,t_a)$ captures  a long memory of the past.
%
%For the particular case $\alpha=1$, from  Eq. (\ref{33eq26}) there exists
%\begin{equation}\label{33eq33}
%   \frac{\partial}{\partial t}G(x, p,t,t_a)=\frac{a^2}{2B_a}\frac{\partial}{\partial x^2} G(x, p,t,t_a)-pU(x)G(x, p,t,t_a),
%\end{equation}
%Using power law waiting time, and taking inverse Laplace transform, there exists
and\cite{Klafter2011First}
\begin{equation}\label{33eq26a01}
  \omega(t_a,t)=\frac{\sin(\pi\alpha)}{\pi}\Big(\frac{t_a}{t}\Big)^\alpha\frac{1}{t_a+t}.
\end{equation}
Especially, if setting $p=0$, then $G(x, p=0, t,t_a)=\int_{0}^{\infty} G(x, A, t,t_a) dA$ reduces to %$G(x, t,t_a)$ 
the distribution of $x$; and Eq. (\ref{33eq24}) turns to the well known master equation of ACTRW model
\begin{equation}\label{33eq34}
\begin{split}
  G(k,p=0,t,t_a) & =\frac{1-\omega(t_a,s)}{s} \\
    & +\frac{1-\phi(s)}{s}\frac{\cos(ka)\omega(t_a,s)}{1-\cos(ka)\phi(s)},
\end{split}
\end{equation}
when $x_0=0$. Eq. (\ref{33eq34}) is a generalization of the Montroll-Weiss
equation for ACTRW. Omitting the motionless part of Eq. (\ref{33eq34}), yields \cite{Barkai2003Aging}
\begin{equation}\label{33eq35}
\begin{split}
_{0}\mathcal{D}_{t}^\alpha G&(x, p=0,t, t_a)=\frac{a^2}{2B_{\alpha}}\frac{\partial^2}{\partial x^2}G(x, p=0,t,t_a)\\
&+1/B_\alpha\omega(t_a,t)*\frac{t^{-\alpha}}{\Gamma(1-\alpha)}\delta(x),
\end{split}
\end{equation}
where $_{0}\mathcal{D}_{ t}^\alpha$ is the Riemann-Liouvile fractional derivative \cite{Metzler:00}, and the notation ``$\ast$" represents the convolution of the functions with respect to $t$. In fact, Eq. (\ref{33eq35}) can also be obtained by taking $p=0$ in Eq. (\ref{33aeq14}).
%is a  special case of the forward Feynman-Kac equation,  which can be obtained directly  from  Eq. (\ref{33aeq14}).
Furthermore, if $t_a=0$, then $\omega(t_a,t)=\phi(t)$, and Eq. (\ref{33eq24}) reduces to the Feynman-Kac equation in frequency domain for the CTRW model \cite{Carmi2010On},
\begin{equation}\label{33aeq24}
\begin{split}
  G(k,p,s,t_a=0) & =\frac{1-\phi(s+pU(-i\frac{\partial}{\partial k}))}{s+pU(-i\frac{\partial}{\partial k})}\\
  &\cdot \frac{Q_0(k,p,s)}{1-\cos(ka)\phi(s+pU(-i\frac{\partial}{\partial k}))}.
\end{split}
\end{equation}
Then by inverse transform,  Eq. (\ref{33aeq24}) yields the imaginary time fractional Schr\"{o}dinger equation \cite{Turgeman2009Fractional}
\begin{equation}\label{33aeq25}
\begin{split}
  \frac{\partial}{\partial t}G(x,p,t) & =\frac{a^2}{2B_\alpha}\frac{\partial^2}{\partial x^2 }D_t^{1-\alpha}G(x,p,t)-pU(x)G(x,p,t).
\end{split}
\end{equation}

\subsection{Continuous step length PDF}
In the following, we consider another case, i.e., the displacement of each step is not a constant but a random variable following a symmetric PDF $f(x)$. It may be Gaussian distribution or symmetrical power law distribution. We consider the first step  of the particle, i.e., the relation between $Q_0(x,A,t)$ and $Q_1(x,A,t,t_a)$. For ACTRW,  there exists
\begin{equation}\label{33eq81}
\begin{split}
  Q_1(x,A,&t,t_a)  =\int_{0}^t\int_{-\infty}^{\infty} \omega(t_a, \tau)f(\Delta x) \\
    & Q_0\Big(x-\Delta x, A-\tau U(x-\Delta x),t-\tau\Big)\\
   & d\Delta xd\tau,
\end{split}
\end{equation}
where the waiting time of the first step is $\omega(t_a, t)$.
For the general case, the variable $A$ may be negative. Performing Fourier transform instead of Laplace transform  to Eq. (\ref{33eq81}), leads to
\begin{equation}\label{33eq82}
\begin{split}
 Q_1(x,p,t,t_a) & =\int_{0}^t\int_{-\infty}^{\infty}\exp( ip\tau U(x-\Delta x)) \omega(t_a, \tau) \\
    & f(\Delta x)Q_0(x-\Delta x, p,t-\tau)d\Delta xd\tau,
\end{split}
\end{equation}
where $p$ is conjugate to $A$.
By Laplace transform with respect to $t$ and Fourier transform with respect to $x$,
\begin{equation}\label{33eq83}
\begin{split}
   Q_1(k,p,s,t_a)&=\int_{-\infty}^{\infty}\int_{-\infty}^{\infty}\exp(ikx)\omega(t_a,s-ipU(x-\Delta x))\\
   &f(\Delta x)Q_0(x-\Delta x,p,s)d\Delta x dx.
\end{split}
\end{equation}
By variable substitution, i.e., $x-\Delta x=y$,
\begin{equation*}%\label{33eq84}
\begin{split}
  Q_1(k,p,s,t_a) &=\int_{-\infty}^{\infty}\int_{-\infty}^{\infty}\exp(iky+ik \Delta x)\\
  &f(\Delta x)\omega(t_a,s-ipU(y))Q_0(y,p,s)d\Delta x dy \\
    & =f(k)\omega\Big(t_a, s-ipU\Big(-i\frac{\partial}{\partial k}\Big)\Big)Q_0(k,p,s).
\end{split}
\end{equation*}
For ACTRW, the waiting times of the other steps, i.e., $n\geq 1$, following a common PDF $\phi(t)$, there exists
\begin{equation*}\label{33eq91}
\begin{split}
  Q_{n+1}&(x,A,t,t_a)=\int_{0}^t\int_{-\infty}^{\infty} \phi( \tau)f(\Delta x)\\
  &Q_n(x-\Delta x, A-\tau U(x-\Delta x),t-\tau,t_a)d\Delta xd\tau.
\end{split}
\end{equation*}
Using double Fourier transform and Laplace transform, we can obtain
\begin{equation*}\label{33eq92}
  Q_{n+1}(k,p,s,t_a)=f(k)\phi\Big(s-ipU\Big(-i\frac{\partial}{\partial k}\Big)\Big)Q_n(k,p,s,t_a).
\end{equation*}
From Eq. (\ref{33eq22}), we have
\begin{equation}\label{33eq93}
\begin{split}
   G(k,p,s,t_a) & =\frac{1-\omega(t_a,s-ipU(-i\frac{\partial}{\partial k}))}{s-ipU(-i\frac{\partial}{\partial k})}\exp(ikx_0) \\
    & +\frac{1-\phi(s-ipU(-i\frac{\partial}{\partial k}))}{s-ipU(-i\frac{\partial}{\partial k})}\\
    &\cdot\frac{f(k)\omega(t_a,s-ipU(-i\frac{\partial}{\partial k}))Q_0(k,p,s)}{1-f(k)\phi(s-ipU(-i\frac{\partial}{\partial k}))}.
\end{split}
\end{equation}
\subsubsection{Power law waiting time}

We first consider the case of power law waiting time and Gaussian displacement.  For simplicity, we ignore the unmoving part of Eq. (\ref{33eq93}).  Using Eqs. (\ref{33eq03}) and (\ref{33eq93}), we have

\begin{equation}\label{33eq94}
\begin{split}
G(k,p,s,t_a)&=\frac{B_\alpha}{(s-ipU(-i\frac{\partial}{\partial k}))^{1-\alpha}} \\
    & \cdot\frac{\omega(t_a, s-ipU(-i\frac{\partial}{\partial k }))}{B_\alpha(s-ipU(-i\frac{\partial}{\partial k}))^\alpha+\frac{k^2}{2}}.
\end{split}
\end{equation}
It can be noticed that Eq. (\ref{33eq94}) can be rearranged by the following formula
\begin{equation*}%\label{33eq95}
\begin{split}
  \Big(\Big(s-ipU&\Big(-i\frac{\partial}{\partial k}\Big)\Big)^\alpha+\frac{k^2}{2B_\alpha}\Big)\Big(s-ipU\Big(-i\frac{\partial}{\partial k}\Big)\Big)^{1-\alpha}\\
  &\cdot G(k,p,s,t_a)=\omega\Big(t_a, s-ipU\Big(-i\frac{\partial}{\partial k}\Big)\Big).
\end{split}
\end{equation*}
Taking inverse Laplace and Fourier  transform  yields
\begin{equation}\label{33eq56}
\begin{split}
  \frac{\partial}{\partial t}G(x,p,t,t_a) & =\frac{1}{2B_\alpha}\frac{\partial^2}{\partial x^2} D_t^{1-\alpha}G(x,p,t,t_a) \\
    & +ipU(x)G(x,p,t,t_a)-\delta(t)\\
    &+\exp(iptU(x))\omega(t_a,t)\delta(x-x_0).
\end{split}
\end{equation}
Furthermore, for the  heavy tailed distribution, i.e.,
\begin{equation}\label{33eq61}
  f(x) \sim \frac{A}{|x|^{1+\beta}}
\end{equation}
with $0<\beta<2$, it has the characteristic function~\cite{Klafter2011First}
%according to the results and $A$ guarantees the normalization, the characteristic function can be shown by
\begin{equation}\label{33eq62}
  f(k)\sim 1-A_\beta|k|^{\beta},
\end{equation}
where $A_\beta=A \pi /(\Gamma(1+\beta)\sin(\pi \beta/2))$. 
%If $\beta>2$ in Eq.~(\ref{33eq61}), there exists $f(k)\sim 1-1/2k^2$.
From Eq. (\ref{33eq62}) and the unmoving part of Eq. (\ref{33eq93}), we obtain the fractional Feynman-Kac equation for the ACTRW,
\begin{equation}\label{33eq63}
\begin{split}
  \frac{\partial}{\partial t}G(x,p,t,t_a) & =\frac{A_\beta}{B_\alpha} \nabla_x^\beta D_t^{1-\alpha}G(x,p,t,t_a) \\
    & +ipU(x)G(x,p,t,t_a)-\delta(t)\\
    &+\exp(iptU(x))\omega(t_a,t)\delta(x-x_0).
\end{split}
\end{equation}
%\begin{equation}\label{33eq63}
%\begin{split}
% D_{t}^{\alpha}G(x,p,t,t_a)&=\frac{A_\beta}{B_\alpha}\nabla_x^\beta G(x,p,t,t_a)\\
% &+\frac{(\exp(ipU(x)t)t^{-\alpha})*\omega(t_a, t)\delta(x)}{\Gamma(1-\alpha)},
%\end{split}
%\end{equation}
where $\nabla_x^\beta$  is the Riesz fractional operator \cite{Metzler20001The,Compte1996Stochastic}, being defined through
\begin{equation}\label{33eq64}
  \mathcal{F}[\nabla_x^\beta f(x)]=-|k|^\beta f(k).
\end{equation}
\subsubsection{Tempered power law waiting time}
Next, we consider the aging effects of Feynman-Kac equation with exponentially tempered power law waiting time probability density \cite{del-Castillo-Negrete2009Truncation,Sokolov2004Fractional,Meerschaert2008Tempered,Bruno2004A}
\begin{equation}\label{33eq71}
  \phi(t)=\ell_{\alpha}(t)\exp(\lambda^\alpha-\lambda t)\sim \frac{1}{-\Gamma(-\alpha)}\exp(-\lambda t)t^{-1-\alpha},
\end{equation}
where $0<\alpha<1$, $\ell_{\alpha}(t)$ is the L\'{e}vy distribution \cite{Krusemann2014First,Metzler20001The} with index $\alpha$, and $\lambda>0$ generally is a small parameter; $\alpha$ controls the power law tail, and $\lambda$ governs the exponential tempering. The introduced tempering forces the  process to converge from non-Gaussian to Gaussian, and the
convergence is very slow, requiring a long time to find the trend. So, with the time passed
by, both the non-Gaussian and Gaussian phenomena can be observed. 
 In Laplace space, stable L\'{e}vy distribution with index $\alpha$ can be shown by $\int_{0}^{\infty}\exp(-st)\ell_{\alpha}(t)=\exp(-s^\alpha)$. Using the Laplace transform with a shift we have
\begin{equation}\label{33eq72}
   \phi(s)=\exp(\lambda^\alpha-(\lambda+s)^\alpha)\sim 1+\lambda^\alpha-(s+\lambda)^\alpha;
\end{equation}
notice that $\phi(0)=1$, which means the PDF is normalized. By tempering, the distribution turns from heavy tails to semi-heavy \cite{Weron2009Heavy}, i.e., the tails are lighter than those of non-Gaussian stables laws, but much heavier than Gaussian distribution, and the existence of conventional moments are ensured, being useful in some applications. Recently, the tempered power law waiting time has been used in many systems; see also more recent works \cite{del-Castillo-Negrete2009Truncation,Rosinski2007Tempering,Sokolov2004Fractional,Meerschaert2008Tempered,Allegrini2003Generalized,Bruno2004A,Deng2016Effects}.
From Eq. (\ref{33eq93}) without the unmoving part, using $f(k)\sim 1-A_\beta\left|k\right|^{\beta}$, we have
%\begin{equation}\label{33eq73}
%\begin{split}
%  G(k,p,s,t_a) & =\frac{(s+\lambda-ipU(-i\frac{\partial}{\partial k}))^\alpha-\lambda^\alpha}{s-ipU(-i\frac{\partial}{\partial k})}\frac{\omega(t_a, s-ipU(-i\frac{\partial}{\partial k}))
%}{(s+\lambda-ipU(-i\frac{\partial}{\partial k})
%)^\alpha-\lambda^\alpha+c|k|^\beta} \\
%    & =\frac{s-ipU(-i\frac{\partial}{\partial k})-\lambda^\alpha(s-ipU(-i\frac{\partial}{\partial k})
%)^{1-\alpha}}{s-ipU(-i\frac{\partial}{\partial k})}\frac{\omega(t_a, s-ipU(-i\frac{\partial}{\partial k}))
%}{s-ipU(-i\frac{\partial}{\partial k})
%+(c|k|^\beta-\lambda^\alpha)(s-ipU(-i\frac{\partial}{\partial k})
%)^{1-\alpha}}
%\end{split}
%\end{equation}
\begin{equation}\label{33eq73}
\begin{split}
  G(k,p,s,t_a) & =\frac{(s+\lambda-ipU(-i\frac{\partial}{\partial k}))^\alpha-\lambda^\alpha}{s-ipU(-i\frac{\partial}{\partial k})}\\
  &\frac{\omega(t_a, s-ipU(-i\frac{\partial}{\partial k}))
}{(s+\lambda-ipU(-i\frac{\partial}{\partial k})
)^\alpha-\lambda^\alpha+A_\beta|k|^\beta}.
\end{split}
\end{equation}
Then  Eq. (\ref{33eq73}) can be given in another way
\begin{widetext}
\begin{equation}\label{33eq74}
\begin{split}
&\frac{(s+\lambda-ipU(-i\frac{\partial}{\partial k}))^\alpha-\lambda^\alpha
+A_\beta\left|k\right|^\beta)(s-ipU(-i\frac{\partial}{\partial k}))}{(s+\lambda-ipU(-i\frac{\partial}{\partial k}))^\alpha-\lambda^\alpha}
G(x,p,t,t_a)=\omega\Big(t_a, s-ipU\Big(-i\frac{\partial}{\partial k}\Big)\Big).
\end{split}
\end{equation}
Taking inverse Laplace and Fourier transforms, we obtain the final result
\begin{equation}\label{33eq75}
\begin{split}
 \frac{\partial}{\partial t}G(x,p,t,t_a) & =A_\beta \nabla^\beta_x (e^{-t(\lambda-ipU(x))}t^{\alpha-1}E_{\alpha,\alpha}(-\lambda^\alpha t^\alpha)\ast \Big(\Big(\frac{\partial}{\partial t}-ipU(x)\Big)G(x,p,t,t_a)-\delta(t)\Big) \\
    &~~~~~~~~~~+e^{iptU(x)}\omega(t_a,t)\delta(x)-\delta(t),
\end{split}
\end{equation}
\end{widetext}
%\begin{equation}\label{33eq75}
%\begin{split}
% &\left(D_{t}^{\alpha,\lambda}-\lambda^\alpha\right)G(x,p,t,t_a)=A_\beta\nabla_x^\beta G(x,p,t,t_a)
% +\exp(iptU(x))\Big[\exp(-\lambda t)~_0\mathcal{D}_{t}^{\alpha}\omega(t_a,t)*1-\lambda^\alpha\omega(t_a,t)*1\Big]\delta(x),
%\end{split}
%\end{equation}
where $\delta(x)$ is the dirac delta function. For tempered power law waiting time, $\omega(t_a,t)$ can be shown through simple integration \cite{Deng2016Effects}.
If $\beta>2$, the operator  $\nabla_x^\beta$ reduces to $\frac{\partial^2}{\partial x^2}$.
%In Laplace space, the operator $ ~_0\mathcal{D}_{t}^{\alpha,\lambda}$ can be shown by $(s+\lambda-ipU(x))^\alpha$; and
Especially, setting $\lambda=0$, Eq. (\ref{33eq75}) reduces to Eq. (\ref{33eq63}). Furthermore, supposing $p=0$, Eq. (\ref{33eq75}) agrees with the aging diffusion equation with tempered power law waiting time and Gaussian step length distribution \cite{Deng2016Effects}.

\section{Derivation of the Backward Feynman-Kac equation from ACTRW}\label{sect22}
We further consider the distribution of $A$, which is  useful in some applications, such as, calculating the first passage time\cite{Redner2001A}, solving the occupation time \cite{Barkai2006Residence,Majumdar2002Local}. One way is to integrate $G(x,p,t)$ over $x$ from $-\infty$ to $\infty$, which is inconvenient in some cases. Letting the process start at $x_0$, we can derive an equation of  $G_{x_0}(A,t,t_a)$, i.e., the backward Feynman-Kac equation for ACTRW. Supposing the length of each step is $a$, and the particles have the same probability to jump to the left or right,  from the definition of ACTRW we have the relation among $G_{x_0}(A,t,t_a)$, $G_{x_0-a}(A,t,t_a)$ and $G_{x_0+a}(A,t,t_a)$
\begin{widetext}
\begin{equation}\label{43eq11}
  G_{x_0}(A,t,t_a)=\int_0^t\omega(t_a,\tau)\frac{1}{2}\Big(G_{x_0+a}(A-\tau U(x_0),t-\tau,t_a)+G_{x_0-a}(A-\tau U(x_0),t-\tau,t_a)\Big)d\tau+\int_t^{\infty}\omega(t_a,\tau)d\tau\delta(A-tU(x_0)).
\end{equation}
Performing Laplace transform with respect to $A$, there exists
\begin{equation}\label{43eq12}
  G_{x_0}(p,t,t_a)=\int_0^t\omega(t_a,\tau)\frac{1}{2}\exp(-p\tau U(x_0))
  \Big(G_{x_0+a}(p,t-\tau,t_a)+G_{x_0-a}(p,t-\tau,t_a)\Big)d\tau+\int_t^{\infty}\omega(t_a,\tau)d\tau \exp(-pt U(x_0)).
\end{equation}
Using Laplace transform, $t\rightarrow s$, we have
\begin{equation}\label{43eq13}
   G_{x_0}(p,s,t_a)=\omega(t_a,s+p U(x_0))\frac{1}{2}
  \Big(G_{x_0+a}(p,s,t_a)+G_{x_0-a}(p,s,t_a)\Big)+\frac{1-\omega(t_a,s+pU(x_0))}{s+pU(x_0)}.
\end{equation}
Taking Fourier transform, $x_0\rightarrow k_0$, leads to
\begin{equation}\label{43eq14}
   G_{k_0}(p,s,t_a)=\omega\Big(t_a,s+p U\Big(-i\frac{\partial}{\partial k_0}\Big)\Big)\cos(k_0 a)
  G_{k_0}(p,s,t_a)+\frac{1-\omega(t_a,s+pU(-i\frac{\partial}{\partial k_0}))}{s+pU(-i\frac{\partial}{\partial k_0})}\delta(k_0).
\end{equation}
Eq. (\ref{43eq14}) can be rewritten as
\begin{equation}\label{43eq21}
\begin{split}
 \Big(s+pU\Big(-i\frac{\partial}{\partial k_0}\Big)\Big) G_{k_0}(p,s,t_a) & = \Big(s+pU\Big(-i\frac{\partial}{\partial k_0}\Big)\Big)\omega\Big(t_a,s+pU\Big(-i\frac{\partial}{\partial k_0}\Big)\Big)\cos(k_0 a)G_{k_0}(p,s,t_a) \\
    & ~~~~+ \Big(1-\omega\Big(t_a,s+pU\Big(-i\frac{\partial}{\partial k_0}\Big)\Big)\Big)\delta(k_0).
\end{split}
\end{equation}
Performing inverse Fourier transform on Eq.~\eqref{43eq21} with respect to $k_0$
\begin{equation}\label{43eq22}
\begin{split}
 (s+pU(x_0)) G_{x_0}(p,s,t_a)-1 & = (s+pU(x_0))\omega(t_a,s+pU(x_0))\Big(1+\frac{a^2}{2}\frac{\partial^2}{\partial x_0^2}\Big)G_{k_0}(p,s,t_a) -\omega(t_a,s+pU(x_0)).
\end{split}
\end{equation}
Further taking inverse Laplace transform with respect to $s$, we get
\begin{equation}\label{43eq23}
\begin{split}
\frac{\partial}{\partial t}G_{x_0}(p,t,t_a) & =\frac{ \partial}{\partial t}\Big(\exp(-ptU(x_0))\omega(t_a,t)\Big(1+\frac{a^2}{2}\frac{\partial^2}{\partial x_0^2}\Big)\Big)*G_{x_0}(p,t,t_a)-pU(x_0)G_{x_0}(p,t,t_a)  \\
    &+pU(x_0)\Big(\exp(-ptU(x_0))\omega(t_a,t)\Big(1+\frac{a^2}{2}\frac{\partial^2}{\partial x_0^2}\Big)\Big)*G_{x_0}(p,t,t_a)-\exp(-ptU(x_0))\omega(t_a,t),
\end{split}
\end{equation}
\end{widetext}
where ``*" is the Laplace convolution operator with respect to $t$. Especially, letting $t_a=0$, $G_{k_0}(p,s,t_a)\rightarrow G_{k_0}(p,s)$ and Eq. (\ref{43eq14}) reduces to the Feynman-Kac equation for CTRW model,
\begin{equation*}
\begin{split}
 G_{k_0}(p,s) & =\phi\Big(s+p U\Big(-i\frac{\partial}{\partial k_0}\Big)\Big)\cos(k_0 a)
  G_{k_0}(p,s) \\
    & +\frac{1-\phi(s+pU(-i\frac{\partial}{\partial k_0}))}{s+pU(-i\frac{\partial}{\partial k_0})}\delta(k_0).
\end{split}
\end{equation*}
Performing inverse transform to the above equation yields \cite{Carmi2010On}
\begin{equation*}
   \frac{\partial}{\partial t}G_{x_0}(p,t)=\frac{a^2}{2B_\alpha}D_{t}^{1-\alpha}\frac{\partial^2}{\partial x_0^2}G_{x_0}(p,t)-pU(x_0)G_{x_0}(p,t).
\end{equation*}
If the functional $A$ is not necessarily positive, see Appendix \ref{Appendix} for its governing equation.

\section{Application}
In this section, we consider some applications of the distribution of the paths of particles performing anomalous diffusion. With the help of functional of anomalous diffusion path, it's convenient to obtain the occupation time in half space, fluctuations of occupation fraction, and the first passage time.
\subsection{Occupation time in half space for ACTRW}
In the following, we consider the occupation time of particles in half space, which is a hot topic in mathematic or physics. We introduce $T^{+}$, the occupation time in $x>0$
\begin{figure}[htb]
  \centering
  % Requires \usepackage{graphicx}
  \includegraphics[width=9cm, height=6cm]{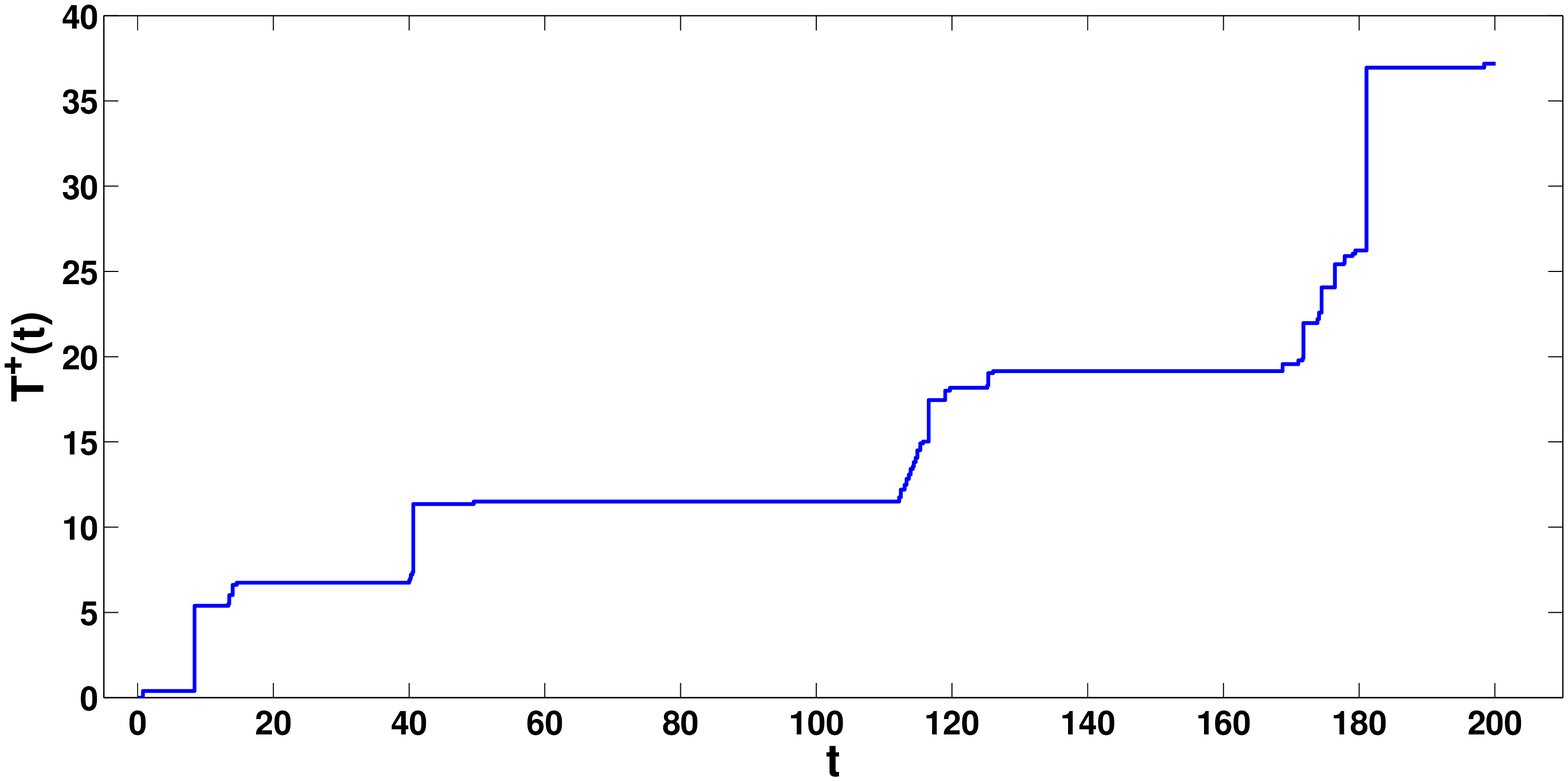}\\
  \caption{Occupation time of an ACTRW with power law waiting time and Gaussian distribution of step lengths (with zero mean and unit dispersion).
}\label{4fig01}
\end{figure}

\begin{figure}[htb]
  \centering
  % Requires \usepackage{graphicx}
  \includegraphics[width=9cm, height=6cm]{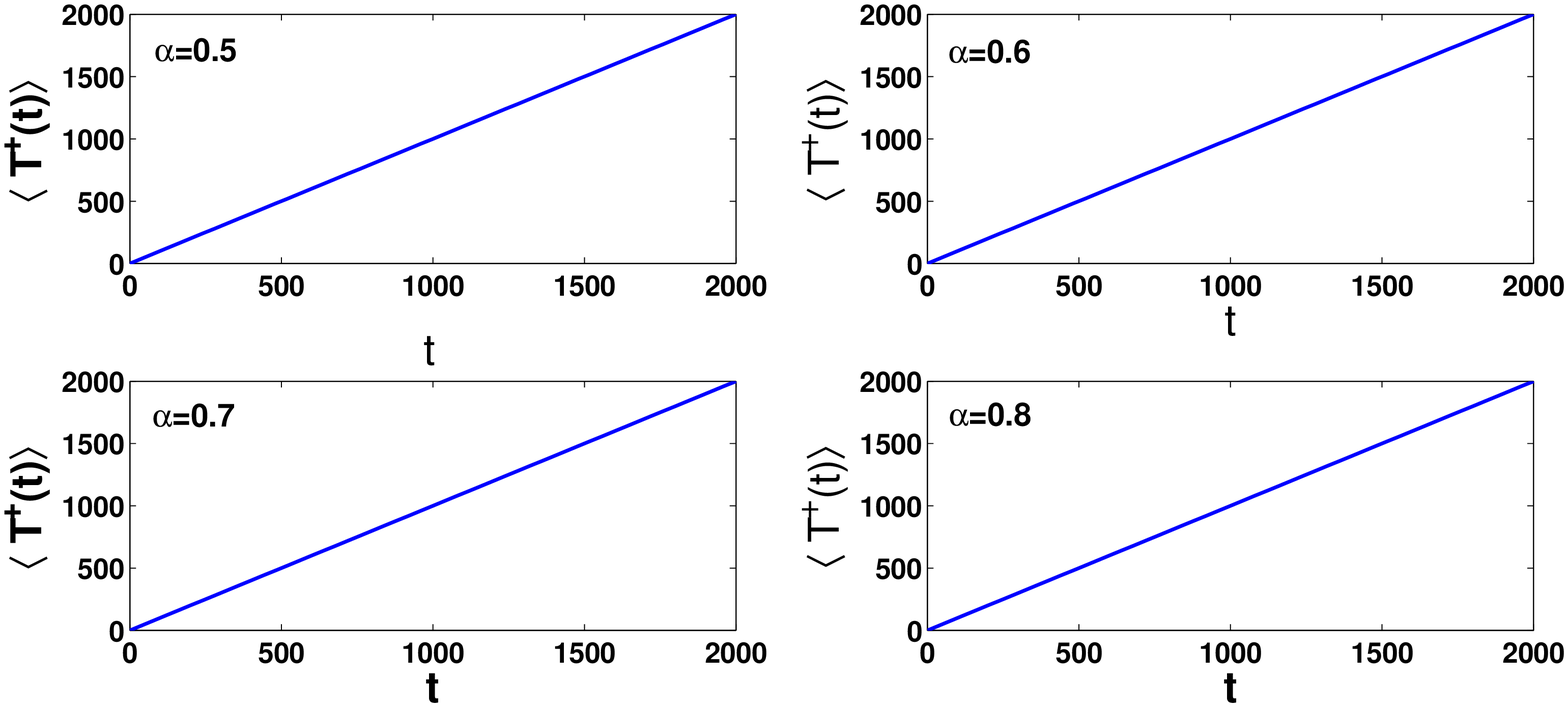}\\
  \caption{ The relation between $\langle T^{+}\rangle$ and the observation time $t$ for various $\alpha$ with $t>t_a$. The number of particles is $4000$, $t_a=10$, $t=2000$, and $\alpha=0.5$, $0.6$, $0.7$, and $0.8$. The real lines are obtained by averaging $4000$ trajectories. It can be noticed that $\langle T^{+}\rangle$ grows linearly with time $t$.
}\label{4fig02}
\end{figure}
\begin{equation}\label{43aeq01}
  T^{+}=\int_0^{t}\theta(x(\tau))d\tau,
\end{equation}
where $\theta(x)=1$ for $x\geq 0$ and is zeros otherwise. In  Fig. \ref{4fig01}, we give a trajectory of $T^{+}$ getting from a trajectory of  ACTRW model. Furthermore, taking $U(x(t))=\theta(x(t))$, we have the relation $A=T^{+}$. In order to derive the PDF of $T^{+}$, we consider the backward Feynman-Kac equation Eq.~(\ref{43eq22}), i.e., power law waiting time and regular jump length, in Laplace space
\begin{equation}\label{43aeq02}
 G_{x_0}(p,s,t_a)=\left\{
  \begin{array}{ll}
    \frac{K_a\omega(t_a,s)}{1-\omega(t_a,s)}\frac{\partial^2}{\partial x_0^{2}}G_{x_0}(p,s,t_a)+\frac{1}{s}, &\hbox{$x_0<0$;} \\
   \frac{ K_a\omega(t_a,s+p)}{1-\omega(t_a,s+p)}\frac{\partial^2}{\partial x_0^{2}}G_{x_0}(p,s,t_a)+\frac{1}{s+p}, &\hbox{$x_0>0$,}
  \end{array}
\right.
\end{equation}
where $K_a=1/2a^2$ and $\omega(t_a,s)$ denotes as the PDF of the forward waiting time.   From Eq. (\ref{33eq26a01}), we obtain %$\omega(t_a,s)$,
\begin{equation}\label{43aeq02a1}
 \omega(t_a,s)=\frac{1}{\Gamma(\alpha)}\exp(s t_a)\Gamma(\alpha,s t_a),
\end{equation}
where $\Gamma(\alpha,y)$ is the incomplete Gamma function \cite{Abramowitz1984Handbook} and $\Gamma(\alpha,y)=\int_y^{\infty}\exp(-z)z^{\alpha-1}dz$. Supposing $G_{x_0}(p,s,t_a)\rightarrow 0$ for $|x_0|\rightarrow \infty$, it's easy to solve the second order, ordinary differential equation about $x_0$,
\begin{equation}\label{43aeq03}
G_{x_0}(p,s,t_a)=\left\{
                   \begin{array}{ll}
                     C_{0}\exp\Big(x_0\sqrt{\frac{1-\omega(t_a,s)}{K_a\omega(t_a,s)}}\Big)+\frac{1}{s}, & \hbox{$x_0<0$;} \\
                    C_{1}\exp\Big(-x_0\sqrt{\frac{1-\omega(t_a,s+p)}{K_a\omega(t_a,s+p)}}\Big)+\frac{1}{s+p}, & \hbox{$x_0>0$.}
                   \end{array}
                 \right.
\end{equation}
Here $C_0$ and $C_1$ are undetermined coefficients. It can be noticed the particles can never reach the right plane if the initial position $x_0\rightarrow -\infty$, i.e., $G_{x_0}(T^{+},t,t_a)=\delta(T^{+})$; furthermore, performing Laplace transform we have $G_{x_0}(p,s,t_a)=\frac{1}{s}$, which  agrees with Eq.~(\ref{43aeq03}). If $x_0\rightarrow +\infty$, the probability of the particles to reach  the left plane is $0$, namely $G_{x_0}(T^{+},t,t_a)=\delta(T^{+}-t)$ and $G_{x_0}(p,s,t_a)=\frac{1}{s+p}$, which is consistent with Eq.~(\ref{43aeq03}). If assuming  that $G_{x_0}(p,s,t_a)$ and its first derivative about $x_0$ are continuous at $x_0=0$, we  obtain

\begin{equation}\label{43eq04}
\left\{
  \begin{array}{ll}
    C_0+\frac{1}{s}=C_{1}+\frac{1}{s+p}& \\
    -C_0\sqrt{\frac{1-\omega(t_a,s)}{K_a\omega(t_a,s)}}=C_1\sqrt{\frac{1-\omega(t_a,s+p)}{K_a\omega(t_a,s+p)}}&.
  \end{array}
\right.
\end{equation}
From the above equation,  there exists
\begin{equation}\label{43beq04}
  \left\{
    \begin{array}{ll}
      C_{0}=-\frac{p}{s(s+p)}\frac{\sqrt{(1-\omega(t_a,s+p))\omega(t_a,s)}}{\sqrt{\omega(t_a,s+p)(1-\omega(t_a,s))}+\sqrt{\omega(t_a,s)(1-\omega(t_a,s+p))}}, &  \\
    C_{1}=\frac{p}{s(s+p)}\frac{\sqrt{(1-\omega(t_a,s))\omega(t_a,s+p)}}{\sqrt{\omega(t_a,s+p)(1-\omega(t_a,s))}+\sqrt{\omega(t_a,s)(1-\omega(t_a,s+p))}}. &
    \end{array}
  \right.
\end{equation}
For simplicity, let the particle start at $x_0=0$, from Eq.~(\ref{43aeq03}) we have $G_0(p,t_a,t)=C_0+1/s$, i.e.,
\begin{widetext}
\begin{equation}\label{43aeq05}
\begin{split}
 G_0(p,s,t_a) =\frac{1}{s(s+p)}\frac{(p+s)\sqrt{(1-\omega(t_a,s))\omega(t_a,s+p)}+s\sqrt{(1-\omega(t_a,s+p))\omega(t_a,s)}}{\sqrt{\omega(t_a,s+p)(1-\omega(t_a,s))}+\sqrt{\omega(t_a,s)(1-\omega(t_a,s+p))}}.
\end{split}
\end{equation}
\end{widetext}
Especially, if $t_a=0$, $\omega(t_a=0,t)=\phi(t)$, we have $\omega(t_a=0,s)=\phi(s)$ and $\omega(t_a=0,s+p)=\phi(s+p)$, and Eq. (\ref{43aeq05}) reduces to
\begin{equation}\label{43aeq06}
  G_0(p,s)=\frac{s^{\alpha/2-1}+(s+p)^{\alpha/2-1}}{s^{\alpha/2}+(s+p)^{\alpha/2}}.
\end{equation}
Using the technique given by Godr\`{e}che and Luck \cite{Godreche2001Statistics}, the PDF of $y=T^{+}/t$ can be shown by the Lamperti PDF \cite{Carmi2010On}, i.e.,
\begin{equation}\label{43aeq07}
\begin{split}
  g(y) &= \frac{\sin(\pi\alpha/2)}{\pi} \\
    &\cdot\frac{y^{\alpha/2-1}(1-y)^{\alpha/2-1}}{y^\alpha+(1-y)^\alpha+2y^{\alpha/2}(1-y)^{\alpha/2}\cos(\pi\alpha/2)}.
\end{split}
\end{equation}
For more details, see \cite{Godreche2001Statistics,Lamperti1958An,Baldassarri1999Statistics}. Eq. (\ref{43aeq05}) works well for all time $t$ and shows the trend of $T^{+}$, while it's difficult to invert Eq.~(\ref{43aeq05}) analytically.
%Hosking method.

\subsection{Fluctuation of occupation fraction}
\begin{figure}[htb]
  \centering
  % Requires \usepackage{graphicx}
  \includegraphics[width=9cm, height=6cm]{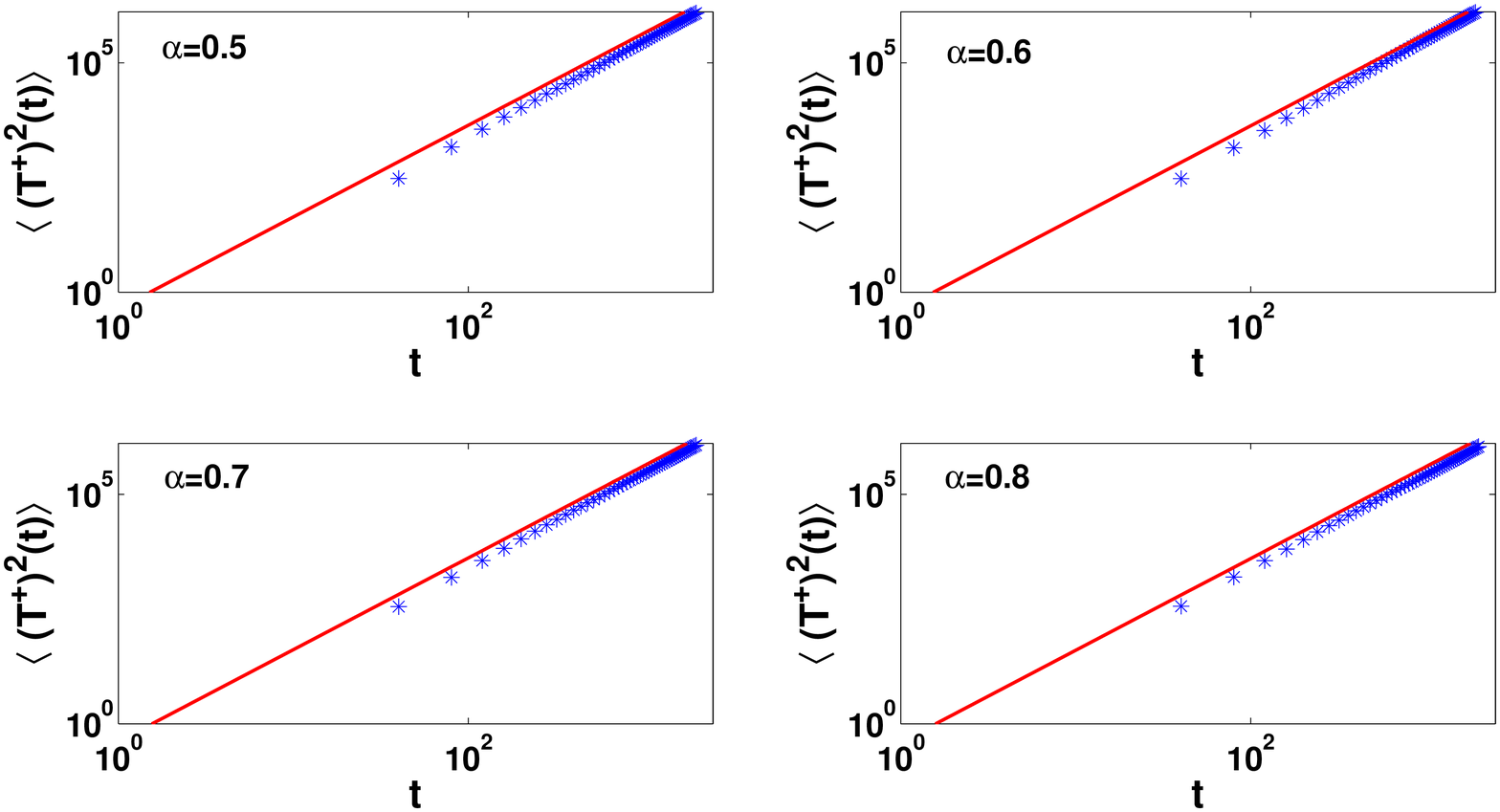}\\
  \caption{ Time evolution of the ensemble average of the occupation time
$\langle (T^+)^2(t)\rangle$ with the waiting time PDF  Eq.~(\ref{33eq01}) for aging slightly. The parameters are taken as $t_a=10$, $t=2000$, and $\alpha=0.5$, $0.6$, $0.7$, and $0.8$. The real (red) lines are for the analytical result Eq.~(\ref{43aeq14}) and the other symbol lines are obtained by averaging $4000$ trajectories.
}\label{4fig03}
\end{figure}

\begin{figure}[htb]
  \centering
  % Requires \usepackage{graphicx}
  \includegraphics[width=9cm, height=6cm]{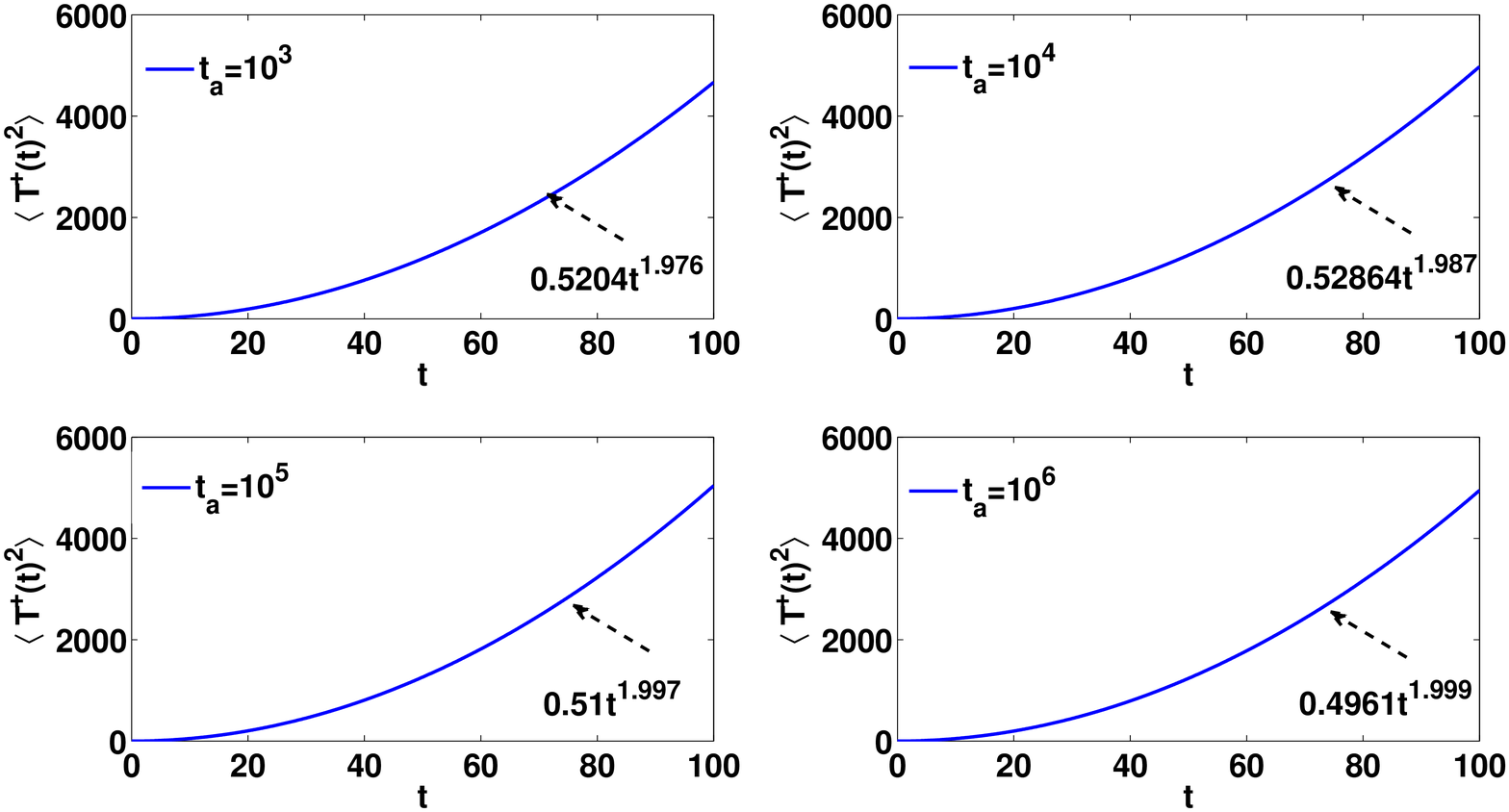}\\
  \caption{ Time evolution of the ensemble average of the occupation time
$\langle (T^+)^2(t)\rangle$ for strongly aging system.
The parameters are taken as $\alpha=0.6$, $t=100$, $N=4000$ and $t_a=10^
3$, $10^4$, $10^5$, and $10^6$. The solid lines are obtained by averaging $4000$ trajectories. $\langle (T^+)^2(t)\rangle=0.5204t^{1.976}$, $0.5286t^{1.987}$, $0.51t^{1.997}$, and $0.4961t^{1.999}$ are the fitting results for small $t$ and large $t_a$, which  agree with Eq.~(\ref{43aeq1503}).
}\label{4fig05}
\end{figure}

\begin{figure}[htb]
  \centering
  % Requires \usepackage{graphicx}
  \includegraphics[width=9cm, height=6cm]{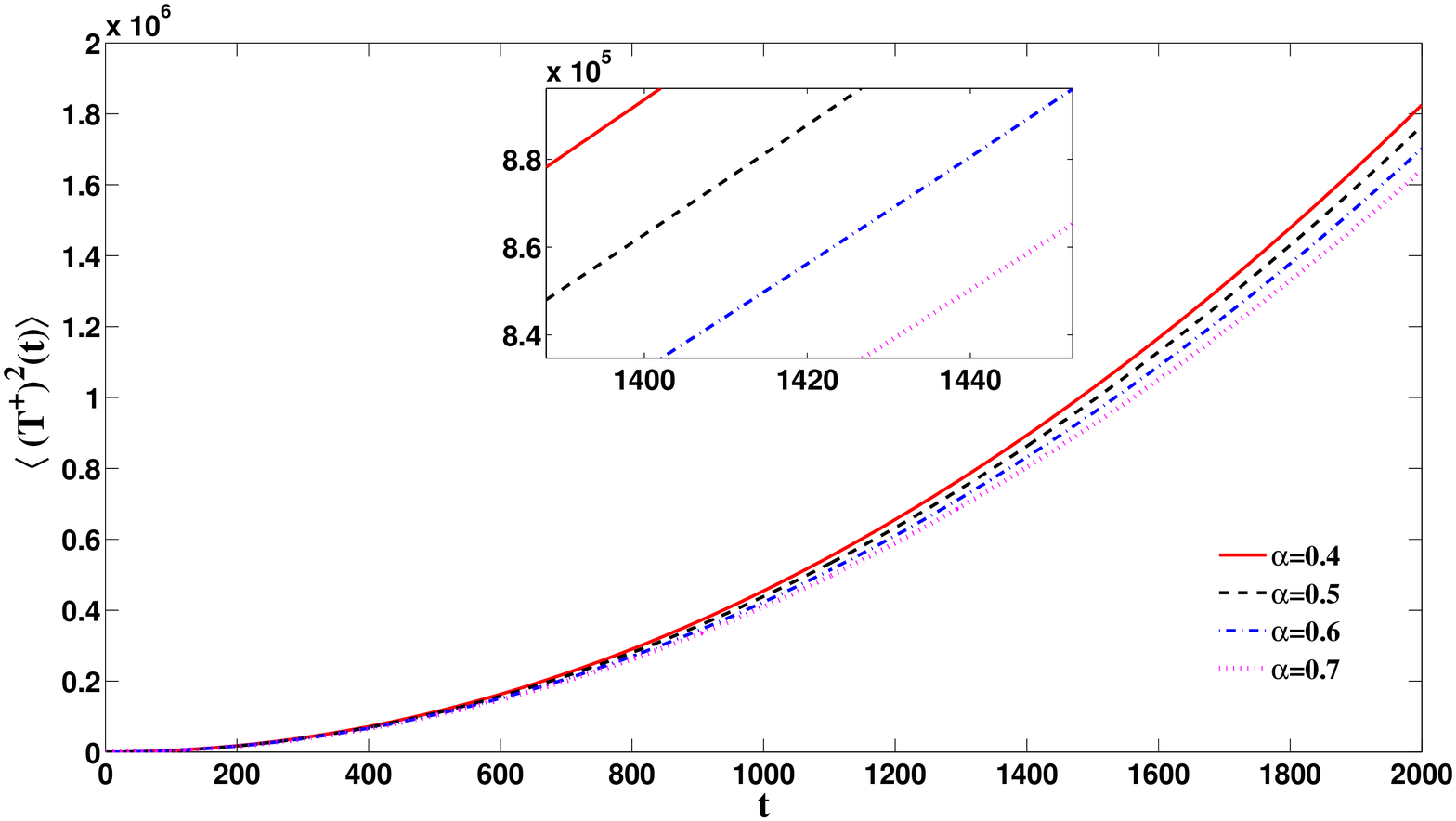}\\
  \caption{ Time evolution of the ensemble average of the occupation time
$\langle (T^+)^2(t)\rangle$ for weak aging system.
The parameters are taken as $t_a=2$, $t=2000$, and $\alpha=0.4$, $0.5$, $0.6$, and $0.7$. The  symbol lines are obtained by averaging $15000$ trajectories.
}\label{4fig04}
\end{figure}

We further introduce $\eta(t)=\frac{T^+}{t}$, a quantity to illustrate the fraction of time that  the particle spends within a given domain\cite{Thaler2002A,Bel2006Weak,Godreche2001Statistics}. %From Eq.~(\ref{43aeq05}), it can be noticed it's not easy to take inverse double Laplace transforms with respect to $p$ and $s$. While, using  the Laplace transform of $G_0(p,s,t_a)$, it's a good choice to obtain the moments of $T^{+}$, by using
Eq.~(\ref{43aeq05}) is very useful to obtain the moments of $T^{+}$, i.e.,
\begin{equation}\label{43aeq11}
  \langle (T^+(s))^n\rangle=(-1)^n \frac{\partial^n}{\partial p^n}G_0(p,s,t_a)|_{p=0}.
\end{equation}
%the existing moments can be obtained by the above equation,  and Eq.~(\ref{43aeq11}) can be obtained by Taylor series.
From Eq.~(\ref{43aeq05}) and Eq.~(\ref{43aeq02a1}), we can obtain
\begin{widetext}
\begin{equation}\label{43aeq11a}
\begin{split}
 G_0(p,s,t_a) =\frac{1}{s(s+p)}\frac{(p+s)\sqrt{(\Gamma(\alpha-e^{st_a}\Gamma(\alpha,st_a))e^{pt_a}\Gamma(\alpha,(s+p)t_a)}+s\sqrt{(\Gamma(\alpha-e^{(s+p)t_a}\Gamma(\alpha,(s+p)t_a))\Gamma(\alpha,st_a)}}{\sqrt{(\Gamma(\alpha-e^{st_a}\Gamma(\alpha,st_a))e^{pt_a}\Gamma(\alpha,(s+p)t_a)}+\sqrt{(\Gamma(\alpha-e^{(s+p)t_a}\Gamma(\alpha,(s+p)t_a))\Gamma(\alpha,st_a)}}.
\end{split}
\end{equation}
\end{widetext}
Since it is difficult to take inverse transform to Eq.~\eqref{43aeq11a} analytically, we just consider the asymptotic behaviors of the moments, inlcuding the first and the second moments of $T^{+}$. % In the following, we discuss the behaviors of the moments of $T^+$. Especially, 
Setting $n=1$ and using the relation between Eq.~(\ref{43aeq11}) and Eq.~(\ref{43aeq11a}), yield
\begin{equation}\label{43aeq145}
\langle T^+(s)\rangle=-\frac{\partial}{\partial p}G_0(p,s,t_a)|_{p=0}=\frac{1}{2s^2},
\end{equation}
by inverse Laplace transform, which leads to
\begin{equation}\label{43aeq1401a}
 \left\{
   \begin{array}{ll}
     \langle T^+(t)\rangle=t/2 & \hbox{} \\
     \langle \eta(t)\rangle=1/2. & \hbox{}
   \end{array}
 \right.
\end{equation}
That is to say, for  both the weakly and strongly aging systems, the aging time $t_a$ makes no difference to $\langle T^+(t)\rangle$. In Fig. \ref{4fig02}, we give the simulation results getting from trajectories for $t\gg t_a$; it can be noticed $\langle T^+(t)\rangle$ increases linearly with time $t$, which  agrees with our theory results Eq.~(\ref{43aeq1401a}). In Laplace space, the second moment of $T^+(t)$ can be given by
\begin{equation}\label{43aeq1401}
\begin{split}
 & \langle T^+(s)^2\rangle  ~ \\
    & =\frac{(st_a)^\alpha-e^{st_a}\Gamma(\alpha,st_a)(4+st_a)\Gamma(\alpha)+4e^{2st_a}\Gamma(\alpha,st_a)^2}{4s^3\Gamma(\alpha,st_a)e^{st_a}(e^{st_a}\Gamma(\alpha,st_a)-\Gamma (\alpha))}.
\end{split}
\end{equation}
We first  consider weakly aging system, i.e., $t_a\ll t$. For small $y$, the incomplete Gamma function has the following relation
\begin{equation}\label{43aeq12}
\begin{split}
  \Gamma(\alpha,y) & =\Gamma(\alpha)\Big(1-\exp(-y)\sum_{n=0}^{\infty}\frac{y^{n+\alpha}}{\Gamma(1+\alpha+n)}\Big) \\
    &\sim \Gamma(\alpha)\Big(1-\exp(-y)\frac{y^\alpha}{\Gamma(1+\alpha)}\Big).
\end{split}
\end{equation}
Using Eq.~(\ref{43aeq1401}) and Eq.~(\ref{43aeq12}), for $t_a\ll t$, i.e., $t_a s\rightarrow 0$, after a few simple calculations, there exists %Eq.~(\ref{43aeq1401}) can be shown by
\begin{equation}\label{43aeq13}
\langle T^+(s)^2\rangle\sim\frac{(1-\alpha/4)}{s^3}.
\end{equation}
Hence, inverse Laplace transforming Eq.~(\ref{43aeq13}) with respect to $s$ yields
\begin{equation}\label{43aeq14}
\langle (T^+)^2(t)\rangle \sim \left(\frac{1}{2}-\frac{\alpha}{8}\right)t^2,
\end{equation}
or $\langle \eta(t)^2\rangle \sim(1/2-\alpha/8)$. In Fig. \ref{4fig03}, it can be noticed the theory results are consistent with the simulation results. In addition, the fluctuation of the occupation fraction is
\begin{equation}\label{43aeq1501}
\begin{split}
  \langle \Delta^2_\eta(t)\rangle & =  \langle \eta^2(t)\rangle-(\langle \eta(t)\rangle)^2 \\
    &\sim \frac{1}{4}-\frac{\alpha}{8}.
\end{split}
\end{equation}
We further consider the strongly aging system, i.e., $t\ll t_a$. For large values of $y$, $\Gamma(\alpha,y)$ behaves as
\begin{equation}\label{43aeq1502}
\Gamma(\alpha,y)\sim y^{\alpha-1}e^{-y}.
\end{equation}
Substituting Eq.~(\ref{43aeq1502}) into Eq.~(\ref{43aeq1401}), there exists
\begin{equation}\label{43aeq1503}
 \langle T^+(t)^2\rangle =\frac{1}{2}t^2.
\end{equation}
Eq.~(\ref{43aeq1503}) is verified by Fig.~\ref{4fig05}. %From the fitting results, we can find that $\langle (T^+)^2(t)\rangle\sim 1/2 t^2$ for strongly aged system.
% from Eq.~(\ref{43aeq05}) there exist
%\begin{widetext}
%\begin{equation}\label{43aeq16}
%  G_0(p,s,t_a)\sim\frac{1}{s(s+p)}\frac{(p+s)\sqrt{(\Gamma(\alpha)-s^{\alpha-1} t_a^{\alpha-1})
%  (s+p)^{\alpha-1}}+s\sqrt{(\Gamma(\alpha)-(s+p)^{\alpha-1} t_a^{\alpha-1})
%  s^{\alpha-1}}}{\sqrt{(\Gamma(\alpha)-s^{\alpha-1} t_a^{\alpha-1})
%  (s+p)^{\alpha-1}}+\sqrt{(\Gamma(\alpha)-(s+p)^{\alpha-1} t_a^{\alpha-1})
%  s^{\alpha-1}}}.
%\end{equation}
%\end{widetext}
%From Eq.~(\ref{43aeq11}), there exists
%\begin{equation}\label{43aeq17}
%\langle T^+(s)\rangle \sim \frac{1}{2s^2},
%\end{equation}
%then we can obtain $\langle T^+(t)\rangle=t/2$, while for the second moments, we have
%\begin{equation}\label{43aeq18}
%\begin{split}
%  \langle (T^+)^2(s)\rangle &\sim \frac{-4s^\alpha t_a^\alpha+(3+\alpha)st_a\Gamma(\alpha)}{-4s^{3+\alpha}t_a^\alpha+4s^4t_a\Gamma(\alpha)} \\
%    &\sim \frac{3+\alpha}{4s^3}.
%\end{split}
%\end{equation}
%Performing Laplace transform, we have
%\begin{equation}\label{43aeq19}
% \langle (T^+)^2(s)\rangle\sim \frac{3+\alpha}{8}t^2,
%\end{equation}
Furthermore, we get
\begin{equation}\label{43aeq20}
  \langle \Delta^2_\eta(t)\rangle\sim\frac{1}{4}.
\end{equation}
From Eq.~(\ref{43aeq20}) and Eq.~(\ref{43aeq1501}), it can be noted that
the fluctuation of occupation fraction for strongly aging system is larger than the one of weakly aging system. This can be intuitively explained as follows. For $t<<t_a$, i.e, the time $t$ is small, the first steps of most  particles have not completed and they still stay at the initial position and long waiting time plays an important role. Furthermore, the coefficient of $\langle (T^+)^2(t) \rangle$ increases as the decrease of $\alpha$ for weakly aging, which is confirmed by Fig.~\ref{4fig04}. While, for strongly aging systems, the coefficient of $\langle (T^+)^2(t) \rangle$ is a constant.
\subsection{First passage time}
First passage times \cite{Bel2005Weak,Buonocore1987A,Krusemann2014First,Weihua2017Mean} are central features of many families of stochastic processes, including Poisson processes, Wiener processes, gamma processes, and Markov chains, to name but a few. The first passage time, also called the first hitting time, is defined as the time $T_f$, that takes a particle starting at $x_0=-b$ to hit $x=0$ for the first time with $b>0$. Using the relation between  the occupation time functional and  the distribution  of the first passage time \cite{Kac1951On}
\begin{equation*}
  P_r(T_f>t)=P_r(\max_{0\leq \tau\leq t} x(\tau)<0)=\lim_{p\rightarrow \infty} G_{x_0}(p,t,t_a),
\end{equation*}
where $G_{x_0}(p,t,t_a)$ is the Laplace transform of $G_{x_0}(T^+,t,t_a)$ w.r.t. $T^+=\int_0^t \theta(x(t))dt$. In fact, $P_r(\displaystyle \max_{0<\tau<t}x(\tau)<0)=P_r(T^+=0)=\lim_{T^+\rightarrow 0}\int_0^{T^+}G_{x_0}(A,t,t_a)dA$; using   the initial value theorem of Laplace transform \cite{Dyke2014An}, there exists
\begin{equation*}
\begin{split}
  P_r(T^+=0) & =\lim_{p\rightarrow \infty}p \mathcal{L}\left[\int_0^{T^+}G_{x_0}(A,t,t_a)dA\right] \\
    & =\lim_{p\rightarrow \infty} G_{x_0}(p,t,t_a).
\end{split}
\end{equation*}
Supposing $x_0=-b$ and $p\rightarrow \infty$, from Eqs.~(\ref{43aeq02}) and  (\ref{43beq04}), we can obtain

\begin{figure}[htb]
  \centering
  % Requires \usepackage{graphicx}
  \includegraphics[width=9cm, height=6cm]{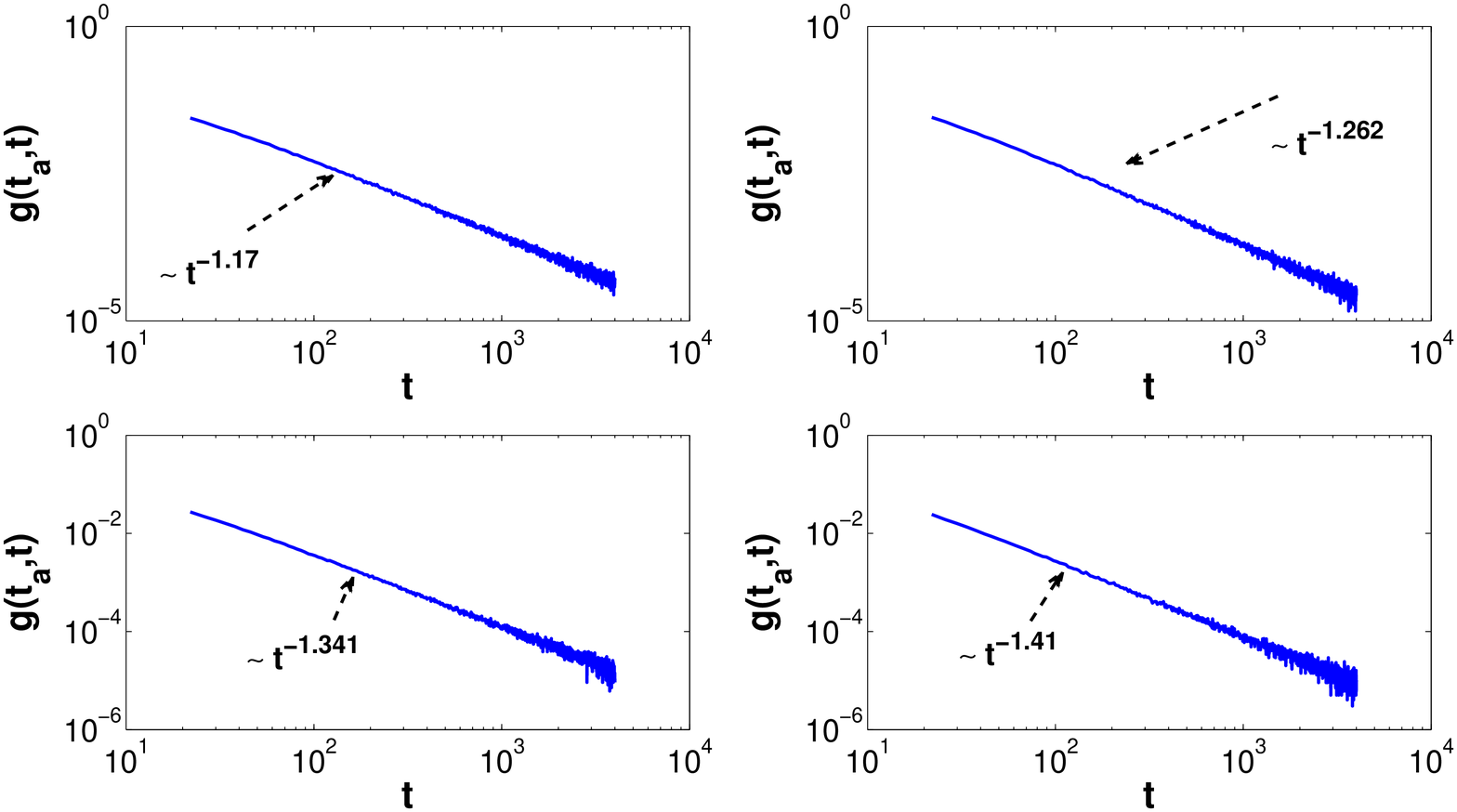}\\
  \caption{ Behaviors of first passage time density $g(t_a,t)$ generated by the trajectories of particles for constant jumping length with $t_a=10$, $t=4000$, $b=0.05$ and $\alpha=0.4$, $0.5$, $0.6$, and $0.7$. The  solid lines are obtained by averaging $10^6$ trajectories and the formulaes are the fitting results.}\label{4fig06}
\end{figure}

\begin{figure}[htb]
  \centering
  % Requires \usepackage{graphicx}
  \includegraphics[width=9cm, height=6cm]{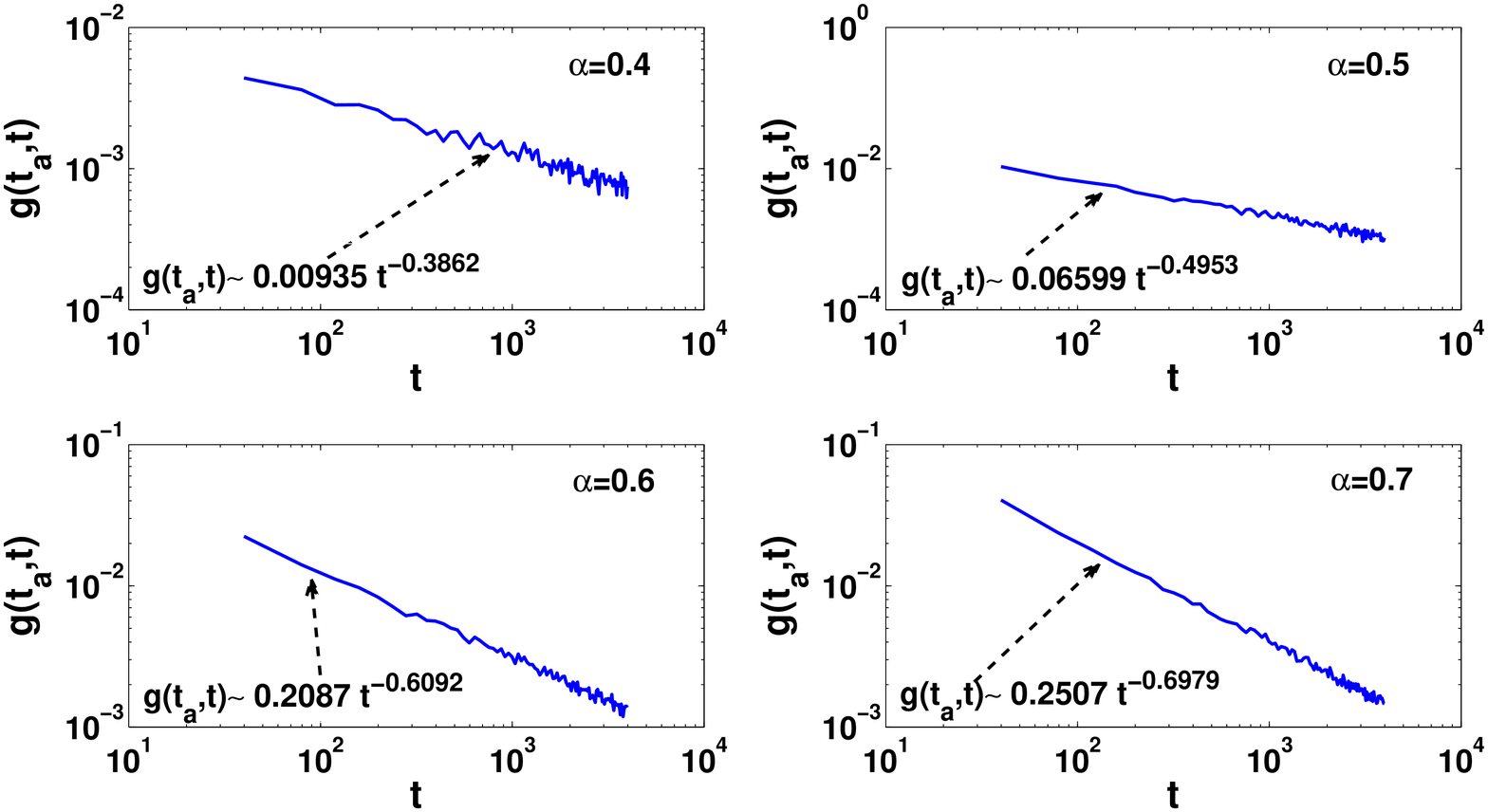}\\
  \caption{ First passage time density for strongly aging system.
The parameters are taken as $t_a=10^5$, $t=4000$, $b=0.05$ and $\alpha=0.4$, $0.5$, $0.6$, and $0.7$. The  solid lines are obtained by averaging $2\times10^5$ trajectories. The formulaes in the lower-left corner of subplots are the fitting results for $\alpha=0.4$, $0.5$, $0.6$, and $0.7$, respectively.
}\label{4fig07}
\end{figure}

\begin{widetext}
\begin{equation}\label{43eqa11}
 \begin{split}
  & \lim_{p\rightarrow \infty}G_{x_0}(p,s,t_a) \\
     & =\lim_{p\rightarrow \infty} \frac{1}{s}-\frac{p}{s(s+p)}\frac{\sqrt{\Gamma(\alpha)-e^{(s+p)t_a}\Gamma(\alpha,(s+p)t_a)\Gamma(\alpha,st_a)}e^{-b\sqrt{\frac{\Gamma(\alpha)-e^{st_a}\Gamma(\alpha,st_a)}{K_{a} e^{st_a}\Gamma(\alpha,st_a)}}}}{\sqrt{\Gamma(\alpha)-e^{(s+p)t_a}\Gamma(\alpha,(s+p)t_a)\Gamma(\alpha,st_a)}+\sqrt{e^{pt_a}\Gamma(\alpha,(s+p)t_a)\Gamma(\alpha)-e^{st_a}\Gamma(\alpha,st_a)\Gamma(\alpha,st_a)}}.
 \end{split}
\end{equation}
\end{widetext}
Using Eq.~(\ref{43aeq1502}), it yields
\begin{equation}\label{43eqa12}
\lim_{p\rightarrow \infty}G_{x_0}(p,s,t_a) =\frac{1}{s}-\frac{1}{s}\exp\Big(-b\sqrt{\frac{\Gamma(\alpha)-e^{st_a}\Gamma(\alpha,st_a)}{K_{a} e^{st_a}\Gamma(\alpha,st_a)}}\Big).
\end{equation}
According to the definition of the first passage time, we can obtain its corresponding PDF
\begin{equation*}%\label{43eqa13}
  g(t_a,t)=\frac{\partial}{\partial t}(1-P_r(T_f>t))=-\frac{\partial}{\partial t}\lim_{p\rightarrow \infty} G_{-b}(p,t,t_a).
\end{equation*}
Performing Laplace transform with respect to $t$, leads to
\begin{equation}\label{43eqa14}
  g(t_a,s)=\exp\left(-b\sqrt{\frac{\Gamma(\alpha)-\exp(st_a)\Gamma(\alpha,st_a)}{K_{a} \exp(st_a)\Gamma(\alpha,st_a)}}\right).
\end{equation}
This demonstrates that Eq.~(\ref{43eqa14}) is valid for all kinds of $s$ and $t_a$. While, it seems difficult to invert Eq.~(\ref{43eqa14}) analytically. In the following, we consider the asymptotic behaviors of the density of the first passage time. For $st_a\rightarrow 0$, i.e., $t_a\ll t$, a simple calculation gives
\begin{equation}\label{43eqa15}
\begin{split}
  g(t_a,s) & =\exp\left(-b\sqrt{\frac{1-e^{st_a}(1-e^{-st_a}\frac{(st_a)^\alpha}{\Gamma(1+\alpha)})}{K_{a} e^{st_a}\left(1-e^{-st_a}\frac{(st_a)^\alpha}{\Gamma(1+\alpha)}\right)}}\right)\\
    &\sim \exp\left(-b\sqrt{\frac{1}{K_{a}\Gamma(1+\alpha)}}(st_a)^{\frac{\alpha}{2}}\right),
\end{split}
\end{equation}
which yields
\begin{equation}\label{43eqa16}
g(t_a,t)\sim\frac{\sqrt{\Gamma(1+\alpha)K_{a}}^{\frac{2}{\alpha}}}{t_ab^{\frac{2}{\alpha}}}\ell_{\frac{\alpha}{2}}\Big(\frac{\sqrt{\Gamma(1+\alpha)K_{a}}^{\frac{2}{\alpha}}}{b^{\frac{2}{\alpha}}}\frac{t}{t_a}\Big),
\end{equation}
with $\ell_\alpha(t)$ being the one sided L\'{e}vy distribution~\cite{Klafter2011First}, and in Laplace space, $\int_0^{\infty} \exp(-st)\ell_\alpha(t)dt=\exp(-s^\alpha)$ for $s>0$. For large values of $\frac{t}{t_a}$, $ g(t_a,t)$ falls off as a power law
\begin{equation}\label{43eqa17}
  g(t_a,t)\sim t_a^{\frac{\alpha}{2}}t^{-1-\frac{\alpha}{2}},
\end{equation}
which is confirmed by Fig.~\ref{4fig06}. For small $t_a$ and large $t$, $ g(t_a,t)\sim t^{-1-\alpha/2}$ and $g(t_a,t)$ tends to $0$ slowly, which is the same as the behaviors of none aging systems\cite{Barkai2001Fractional}.
\begin{figure}[htb]
  \centering
  % Requires \usepackage{graphicx}
  \includegraphics[width=9cm, height=6cm]{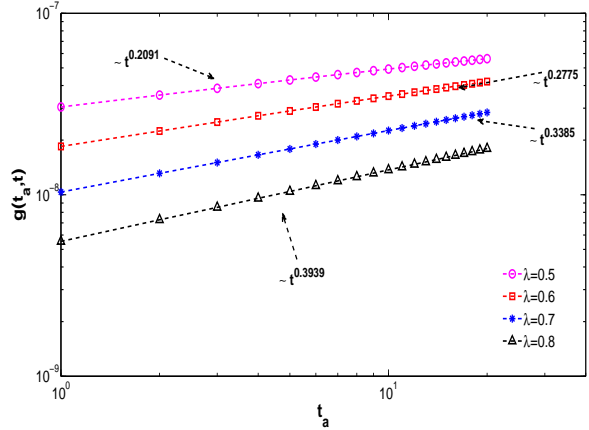}\\
  \caption{The relation between first passage time density and $t_a$ for weakly aging system.
The parameters are taken as $t_a=20$, $t=10^6$, $b=0.05$ and $\alpha=0.5$, $0.6$, $0.7$, and $0.8$. The symbol lines are obtained by the inversion of Eq.~ (\ref{43eqa15}). The formulaes  are the fitting results for $\alpha=0.5$, $0.6$, $0.7$, and $0.8$, respectively.
}\label{4fig08}
\end{figure}
While, if $st_a\rightarrow \infty$, corresponding to $t\ll t_a$, from Eq.~ (\ref{43eqa14}), we have
\begin{equation}\label{43eqa18}
g(t_a, s)\sim \exp\left(-b\sqrt{\frac{\Gamma(\alpha)}{K_{a} t_a^{\alpha-1}}}s^{\frac{1-\alpha}{2}}\right).
\end{equation}
Taking the inversion of Eq.~(\ref{43eqa18}), there exists
\begin{equation}\label{43eqa19}
  g(t_a,t)\sim\Big(\Big(\frac{\sqrt{K_{a}}}{b\sqrt{\Gamma(\alpha)}}\Big)^{\frac{2}{1-\alpha}}/t_a\Big)\ell_{\frac{1-\alpha}{2}}\Big(\Big(\frac{\sqrt{K_{a}}}{b\sqrt{\Gamma(\alpha)}}\Big)^{\frac{2}{1-\alpha}} \frac{t}{t_a}\Big).
\end{equation}
Keeping in mind that $t/t_a$ is a small number, therefore, we can not expand $\ell_\alpha(t)$ as $t^{-\alpha-1}$ to obtain its asymptotic form. 
%In order to obtain its asymptotic behavior, Eq.~(\ref{43eqa18}) is done numerically, and we can get $g(t_a,t)\sim t^{-\alpha}$.
By numerically inverting Eq.~\eqref{43eqa18} and combining with the form of Eq.~\eqref{43eqa19}, we predict that
%
%Using relation between $t/t_a$ and Eq.~(\ref{43eqa19}), there exists
\begin{equation}\label{43eqa20}
g(t_a,t)\sim t_a^{\alpha-1}t^{-\alpha}.
\end{equation}
Eq.~(\ref{43eqa20}) is confirmed by the simulation of trajectories of  the particles; see Fig.~\ref{4fig07}.
%\begin{figure}[htb]
%  \centering
%  % Requires \usepackage{graphicx}
%  \includegraphics[width=9cm, height=6cm]{fig2.eps}\\
%  \caption{ First passage time density with the relation $t$ for strong aging.
%The parameters are taken as $t_a=10^5$, $t=4000$, $b=0.05$ and $\alpha=0.4$, $0.5$, $0.6$, and $0.7$. The  solid lines are obtained by averaging $2*10^5$ trajectories. The formulaes in the lower-left corner of subplots are the fitting results for $\alpha=0.4$, $0.5$, $0.6$, and $0.7$, respectively.
%}\label{4fig07}
%\end{figure}
From Eq.~(\ref{43eqa17}) and (\ref{43eqa20}), we can find that the aging times play an important role for the first passage time. For the weakly aging system, the scaling exponent of $t$ is $-1-\frac{\alpha}{2}$, which changes to $-\alpha$ for strong aging system. While, for the weakly aging system, the power of $t_a$ is $\alpha$, which becomes $\alpha-1$ for strongly aging system.
%While, Eq.~(\ref{43eqa17}) is not valid for $t_a\rightarrow 0 $.

\section{SUMMARY AND DISCUSSION}\label{sect4}
The functionals of the trajectories of particles are very general statistical observables characterizing the motion of particles. In this  paper, based on the  ACTRW, we derive the forward and backward Feynman-Kac equations governing the distribution of the functionals of the trajectories of particles performing anomalous diffusion with (tempered) power law waiting time and/or jump length distributions. For deep studying the aging phenomena of anomalous diffusion, according to the built models, we more specifically calculate the statistical observables: fraction of the occupation time; first passage time. The fluctuation of the occupation fraction is also analyzed, discovering that the aging time has no influence on its first moment but greatly impacts the second moment. Another striking discovery is for the distribution of the first passage time $g(t_a,t)$, which shows that for slightly (none) aging systems, $g(t_a,t)\sim t_a^{\frac{\alpha}{2}}t^{-1-\frac{\alpha}{2}}$, while for strongly aging systems, $g(t_a,t)\sim t_a^{\alpha-1}t^{-\alpha}$.
%
%
%
%In Sect.~(\ref{sect3}), we consider the forward and backward Feynman-Kac equations with the functional $A=\int_0^tU(x(\tau),\tau)d\tau$ based on CTRW model. For simplify, we consider the step length of particles is a constant $a$ and power law waiting time with the index of $0<\alpha<1$, then we discuss the case of Gaussian PDF of length step, finally, the heavy tailed transition length is studied.
%
%
%In the end, we given a application of Feynman-Kac equation.
%
%Recently, the first passage time and the fluctuation of occupation fraction have been studied in the context of dynamical systems, it remain to be seen whether we can obtain them from Feynamn-Kac equation. Furthermore, supposing  $U(x(t))$ is a function of $x(t)$ and $t$, i.e., $U(x(t),t)$,   are there any big difference? they maybe  interesting issues.

\section*{Acknowledgments} 
%The authors thank Eli Barkai for the discussions.
This work was supported by
%the Fundamental Research Funds for the Central Universities under Grant No. lzujbky-2015-77, and
the National Natural Science Foundation of China under Grant No. 11671182, and the Fundamental Research Funds for the Central Universities under Grant No. lzujbky-2017-it57 and lzujbky-2017-ot10.

\appendix
%\appendixpage
%\addcontentsline{toc}{chapter}{¸½Â¼}
%\markboth{¸½Â¼}{}
\begin{appendices}
\section{Backward Feynman-Kac equation with power law jump length } \label{Appendix}
Here we derive the backward Feynman-Kac equation, and the functional is not necessary positive. From the definition of ACTRW model, the particle waits a time $\tau$ at the position of $x_0$ with the waiting time PDF $\omega(t_a,\tau)$, then moves to $x_0+\Delta x$, the PDF of $\Delta x$ being  $f(\Delta x)$. For simplicity, supposing $f(\Delta x)$ has the asymptotic behavior of power law, i.e., $f(\Delta x)\sim |\Delta x|^{-\beta-1}$, there exists
\begin{equation}\label{34eq01}
\begin{split}
   G_{x_0}(A,t,t_a) & =\int_0^t\omega(t_a,t)\int_{-\infty}^{\infty}f(\Delta x)\\
    &  G_{x_0+\Delta x}(A-\tau U(x_0),t-\tau,t_a)d\Delta x d\tau\\
    &+\Big(1-\int_0^t\omega(t_a,\tau)d\tau\Big)\delta(A-tU(x_0)).
\end{split}
\end{equation}
Performing Fourier transform with respect to $A$ leads to
\begin{widetext}
\begin{equation}\label{34eq02}
\begin{split}
   G_{x_0}(p,t,t_a)   =\int_0^t\omega(t_a,t)\int_{-\infty}^{\infty}\exp(iptU(x_0))f(\Delta x)
     G_{x_0+\Delta x}(p,t-\tau,t_a)d\Delta x d\tau
    +\Big(1-\int_0^t\omega(t_a,\tau)d\tau\Big)\exp(iptU(x_0)).
\end{split}
\end{equation}
Then by Laplace and Fourier transform, we have
\begin{equation}\label{34eq03}
\begin{split}
  G_{k_0}(p,s,t_a) & =\omega\Big(t_a,s-ipU\Big(-i\frac{\partial}{\partial k_0}\Big)\Big)f(k_0)G_{k_0}(p,s,t_a)
    +\frac{1-\omega(t_a,s-ipU(-i\frac{\partial}{\partial k_0}))}{s-ipU(-i\frac{\partial}{\partial k_0})}\delta(k_0).
\end{split}
\end{equation}
Rearranging Eq.~(\ref{34eq03}), yields
\begin{equation}\label{34eq04}
\begin{split}
  \Big(s-ipU\Big(-i\frac{\partial}{\partial k_0}\Big)\Big)G_{k_0}(p,s,t_a) & =\Big(s-ipU\Big(-i\frac{\partial}{\partial k_0}\Big)\Big)\omega\Big(t_a,s-ipU\Big(-i\frac{\partial}{\partial k_0}\Big)\Big)f(k_0)G_{k_0}(p,s,t_a) \\
    & +\Big(1-\omega\Big(t_a,s-ipU\Big(-i\frac{\partial}{\partial k_0}\Big)\Big)\Big)\delta(k_0).
\end{split}
\end{equation}
Performing inverse transform, we obtain the final result
\begin{equation}\label{34eq05}
\begin{split}
 \frac{\partial}{\partial t}G_{x_0}(p,t,t_a) =& \frac{\partial}{\partial t}\Big(\exp(iptU(x_0))\omega(t_a,t)\Big)*\Big(1+A_\beta \nabla_{x_0}^\beta\Big)G_{x_0}(p,t,t_a)+ipU(x_0)G_{x_0}(p,t,t_a) \\
    &-ipU(x_0)\Big(\exp(iptU(x_0))\omega(t_a,t)\Big)*\Big(1+A_\beta \nabla_{x_0}^\beta\Big)G_{x_0}(p,t,t_a)-\exp(iptU(x_0))\omega(t_a,t),
\end{split}
\end{equation}
where $\nabla^\beta_{x_0}$ is the Riesz operator, in Fourier space, defined as, $\mathcal{F}[\nabla_x^{\beta} g(x)]=-|k|^\beta g(k)$; compared to Eq.~\eqref{43eq23}, Eq.~\eqref{34eq05} is a more general result.
In particular, if $\beta=2$, there exists $\nabla_{x_0}^{\beta}=\frac{\partial^2}{\partial x_{0}^2}$.
\end{widetext}

\end{appendices}
%
%
%
%
%
%
%
%
%
%
%
%
%
%

%\newpage %Just because of unusual number of tables stacked at end
%\begin{thebibliography}{10}
%%\bibliography{apssamp}% Produces the bibliography via BibTeX.
%
%\bibitem{Carmi:01}  S. Carmi, L. Turgeman£¬E. Barkai, J. Stat. Phys. {\bf141}, 1071 (2010).
%
%\bibitem{Barkai:5}E. Barkai, Phys. Rev. Lett. 90, 104101 (2003).
%\bibitem{Negrete:1} D. del-Castillo-Negrete, Phys. Rev. E {\bf 79}, 031120 (2009).
%
%\bibitem{Rosinski:1} J. Rosi\'{n}ski, Stochastic Process. Appl. {\bf 117}, 677-707 (2007).
%\v{\'{}}
%\bibitem{Stanislavsky:1} Stanislavsky, A.,  Weron, K., Weron, A.: Diffusion and relaxation controlled by tempered $\alpha$-stable processes. Phys. Rev. E {\bf 78}, 051106 (2008).
%
%\bibitem{Sokolov:1} I.M., Sokolov, A.V. Chechkin, J. Klafter, Phys. A {\bf336}, 245-251 (2004)
%
%\bibitem{Meerschaert:2} M.M. Meerschaert, Yong Zhang, B. Baeumer, Geophys. Res. Lett. {\bf35}, L17403 (2008).
%
%
%
%\bibitem{Allegrini:1} P. Allegrini, G. Aquino, P. Grigolini, L., Palatella, A., Rosa, Phys. Rev. E {\bf 68}, 056123 (2003).
%
%\bibitem{Bruno:1} R. Bruno, L. Sorriso-Valvo, V. Carbone, B. Bavassano,  Europhys. Lett. {\bf 66}, 146-152 (2004).
%
%\bibitem{Luck:01}  C. Godr\`{e}che, J.M. Luck, J. Stat. Phys. {\bf104}, 489-524 (2001).
%%
%%
%\end{thebibliography}
                                                                               .

%\bibliographystyle{prestyle}
%\bibliography{wenxian}

\end{document}